\documentclass[onecolumn]{IEEEtran}
\usepackage{amsthm, amsmath, amssymb, amsfonts, url, booktabs, tikz, setspace, fancyhdr, bm, bbm}
\usepackage{multirow}
\usepackage{booktabs}
\usepackage{algorithmic}
\usepackage[ruled,linesnumbered]{algorithm2e}
\usepackage{array}
\usepackage{arydshln}
\usepackage[justification=centering,caption=false,font=normalsize,labelfont=sf,textfont=sf]{subfig}
\usepackage{textcomp}
\usepackage{stfloats}
\usepackage{comment}
\usepackage{verbatim}
\usepackage{graphicx}
\usepackage[noadjust]{cite}
\usepackage{xcolor}
\usetikzlibrary{matrix}
\usepackage{hyperref}
\usepackage{cleveref}

\newtheorem{theorem}{Theorem}
\newtheorem{definition}{Definition}
\newtheorem{lemma}{Lemma}
\newtheorem{claim}{Claim}

\newtheorem{remark}{Remark}

\newtheorem{corollary}{Corollary}

\newtheorem{problem}{Problem}

\begin{document}

\title{Reconstruction Codes for Deletions and Insertions: Connection, Distinction, and Construction}

\author{Yubo~Sun and Gennian~Ge%
\thanks{This research was supported by the National Key Research and Development Program of China under Grant 2020YFA0712100, the National Natural Science Foundation of China under Grant 12231014, and Beijing Scholars Program.}
\thanks{Y. Sun ({\tt 2200502135@cnu.edu.cn}) and G. Ge ({\tt gnge@zju.edu.cn}) are with the School of Mathematical Sciences, Capital Normal University, Beijing 100048, China.}
}

\maketitle

\begin{abstract}
    Let $\mathcal{B}(\cdot)$ be an error ball function.
    A set of $q$-ary sequences of length $n$ is referred to as an \emph{$(n,q,N;\mathcal{B})$-reconstruction code} if each sequence $\bm{x}$ within this set can be uniquely reconstructed from any $N$ distinct elements within its error ball $\mathcal{B}(\bm{x})$. 
    The main objective in this area is to determine or establish bounds for the minimum redundancy of $(n,q,N;\mathcal{B})$-reconstruction codes, denoted by $\rho(n,q,N;\mathcal{B})$.
    In this paper, we investigate reconstruction codes where the error ball corresponds to either the \emph{$t$-deletion ball} $\mathcal{D}_t(\cdot)$ or the \emph{$t$-insertion ball} $\mathcal{I}_t(\cdot)$.
    Our primary technical contributions include:
    \begin{itemize}
      \item Establishing a fundamental connection between reconstruction codes for deletions and insertions. 
      Specifically, for any positive integers $n,t,q,N$, any $(n,q,N;\mathcal{I}_t)$-reconstruction code is also an $(n,q,N;\mathcal{D}_t)$-reconstruction code.
      This leads to the inequality $\rho(n,q,N;\mathcal{D}_t)\leq \rho(n,q,N;\mathcal{I}_t)$.
      \item Identifying a significant distinction between reconstruction codes for deletions and insertions when $N=O(n^{t-1})$ and $t\geq 2$.
          For deletions, we prove that $\rho(n,q,\tfrac{2(q-1)^{t-1}}{q^{t-1}(t-1)!}n^{t-1}+O(n^{t-2});\mathcal{D}_t)=O(1)$, which disproves a conjecture posed in \cite{Chrisnata-22-IT}.
          In contrast, for insertions, we show that $\rho(n,q,\tfrac{(q-1)^{t-1}}{(t-1)!}n^{t-1}+O(n^{t-2});\mathcal{I}_t)=\log\log n + O(1)$, which extends a key result from \cite{Ye-23-IT}.
        
      \item Constructing $(n,q,N;\mathcal{B})$-reconstruction codes, where $\mathcal{B}\in \{\mathcal{D}_2,\mathcal{I}_2\}$, for $N \in \{2,3, 4, 5\}$ and establishing respective upper bounds of $3\log n+O(\log\log n)$, $3\log n+O(1)$, $2\log n+O(\log\log n)$ and $\log n+O(\log\log n)$ on the minimum redundancy $\rho(n,q,N;\mathcal{B})$.
      This generalizes results previously established in \cite{Sun-23-IT}.
    \end{itemize}
\end{abstract}


\section{Introduction}
Motivated by certain emerging storage media, such as DNA storage \cite{Church-12-science, Goldman-13-Nature} and racetrack memory \cite{Chee-18-IT}, which can provide multiple noisy reads, the problem of \emph{reconstruction codes} was introduced by Cai et al. \cite{Cai-22-IT} and Chrisnata et al. \cite{Chrisnata-22-IT}.
Let $\Sigma_q^n$ denote the set of all $q$-ary sequences of length $n$, and let $\mathcal{B}(\cdot)$ represent an error ball function. An \emph{$(n,q,N;\mathcal{B})$-reconstruction code} is defined as a code $\mathcal{C} \subseteq \Sigma_q^n$ such that for any two distinct codewords $\bm{x}, \bm{y} \in \mathcal{C}$, the intersection of their error balls satisfies $|\mathcal{B}(\bm{x}) \cap \mathcal{B}(\bm{y})| < N$. In essence, an $(n,q,N;\mathcal{B})$-reconstruction code guarantees the reconstruction of each codeword $\bm{x}$ from any $N$ distinct elements within its error ball $\mathcal{B}(\bm{x})$. When $N = 1$, reconstruction codes are equivalent to standard error-correcting codes. Therefore, a natural assumption in the context of reconstruction codes is that $N \geq 2$.

In this paper, we investigate the problem of determining or establishing bounds on the minimum redundancy of $(n,q,N;\mathcal{B})$-reconstruction codes, denoted by $\rho(n,q,N;\mathcal{B})$. Our focus lies on the scenarios where the error ball $\mathcal{B}(\cdot)$ represents either the \emph{$t$-deletion ball} $\mathcal{D}_t(\cdot)$ or the \emph{$t$-insertion ball} $\mathcal{I}_t(\cdot)$. According to existing literature, there are primarily three main approaches to investigate the minimum redundancies, $\rho(n,q,N;\mathcal{D}_t)$ and $\rho(n,q,N;\mathcal{I}_t)$.

The first approach employs graph-theoretic tools to analyze theoretical bounds. The most notable work in this area is by Chrisnata et al. \cite{Chrisnata-22-IT}, who established clique cover bounds for $(n,q,2;\mathcal{B})$-reconstruction codes, where $\mathcal{B}\in \{\mathcal{D}_1,\mathcal{I}_1\}$, and demonstrated that 
\begin{equation}\label{eq:lb}
    \rho(n,q,2;\mathcal{B})\leq \log\log n+O(1).
\end{equation}
This method has wide applications, for example, Sun et al. \cite{Sun-23-IT-burst} used it to derive bounds for single burst-insertion/deletion reconstruction codes. Additionally, Banerjee et al. \cite{Banerjee-24-IT} and Sun and Ge \cite{Sun-25-IT-C} generalized this approach to establish bounds for error-correcting codes in nanopore sequencing, and Lu et al. \cite{Lu-24-ITW} employed this method to provide bounds for synthesis-defect correcting codes. However, whether this approach can be generalized to provide lower bounds for insertion/deletion reconstruction codes in other parameter settings remains open.

The second approach utilizes algebraic tools to construct reconstruction codes, thereby establishing upper bounds. 
For the case where $ t=1 $, asymptotically optimal $(n,q,N;\mathcal{B})$-reconstruction codes, where $\mathcal{B}\in \{\mathcal{D}_1,\mathcal{I}_1\}$, have been proposed in the literature \cite{Cai-22-IT, Sun-25-IT-C, Sun-25-arXiv, Chee-18-IT}.
In particular, combining with Equation (\ref{eq:lb}), Cai et al. \cite{Cai-22-IT} showed that
\begin{equation}\label{eq:t=1}
    \rho(n,q,N;\mathcal{B})=
      \begin{cases}
        \log \log n+O(1), & \mbox{if } N=2; \\
        0, & \mbox{if } N\geq 3.
      \end{cases}
\end{equation}
However, the case where $ t \geq 2 $ has largely remained unexplored and previous work has primarily focused on the binary alphabets. 
For deletions, Chrisnata et al. \cite{Chrisnata-22-IT} constructed $(n,2,5;\mathcal{D}_2)$-reconstruction codes with $2 \log n + O(\log \log n)$ bits of redundancy. Subsequently, Sun and Ge \cite{Sun-23-IT} improved their construction and developed codes for $N\in \{2,3,4,5\}$. Specifically, they showed that 
\begin{equation}\label{eq:t=2}
  \rho(n,2,N;\mathcal{D}_2)\leq 
  \begin{cases}
    3\log n+O(\log\log n), & \mbox{if } N=2; \\
    3\log n+O(1), & \mbox{if } N=3;\\
    2\log n+O(\log \log n), & \mbox{if } N=4; \\
    \log n+\log\log n+O(1), & \mbox{if } N\geq 5.
  \end{cases}
\end{equation}
For insertions, Ye et al. \cite{Ye-23-IT} provided $(n,2,5;\mathcal{I}_2)$-reconstruction codes with $\log n + O(\log \log n)$ bits of redundancy.

The third approach examines the maximum intersection size between error balls centered at two distinct sequences selected from a given code $\mathcal{C}\subseteq \Sigma_q^n$.
This direction aligns with Levenshtein's sequence reconstruction problem, which has been extensively studied in the references \cite{Levenshtein-01-IT, Levenshtein-01-JCTA, Sala-17-IT, Gabrys-18-IT, Chrisnata-22-IT, Ye-23-IT, Pham-25-JCTA}.
Let $\nu(\mathcal{C},\mathcal{B})\triangleq \max_{\bm{x}\neq\bm{y}\in \mathcal{C}} \{|\mathcal{B}(\bm{x})\cap \mathcal{B}(\bm{y})| \}$, then $\mathcal{C}$ is an $(n,q,\nu(\mathcal{C},\mathcal{B})+1;\mathcal{B})$-reconstruction code.
\begin{itemize}
    \item When $\mathcal{C}$ is the entire space $\Sigma_q^n$, Levenshtein \cite{Levenshtein-01-JCTA} completely determined $\nu(\Sigma_q^n,\mathcal{D}_t)$ and $\nu(\Sigma_q^n,\mathcal{I}_t)$:
        \begin{gather*}
          \nu(\Sigma_q^n,\mathcal{D}_t)= \frac{2}{(t-1)!}n^{t-1}+O(n^{t-2}),\\
          \nu(\Sigma_q^n,\mathcal{I}_t)= \frac{2(q-1)^{t-1}}{(t-1)!}n^{t-1}+O(n^{t-2}).
        \end{gather*}
    \item When $\mathcal{C}$ is an $(\ell-1)$-deletion correcting code, previous work has primarily focused on the binary alphabets. 
    Gabrys and Yaakobi \cite{Gabrys-18-IT} calculated the quantity $\nu(\mathcal{C},\mathcal{D}_t)$ for the special case where $t\geq \ell=1$.
    Recently, for arbitrary $t\geq \ell\geq 1$, Pham et al. \cite{Pham-25-JCTA} proved that
    \begin{gather}\label{eq:Pham}
        \nu(\mathcal{C},\mathcal{D}_t)\leq \frac{\binom{2\ell}{\ell}}{(t-\ell)!}n^{t-\ell}+O(n^{t-\ell-1}).
    \end{gather}
    \item When $\mathcal{C}$ is an $(\ell-1)$-insertion correcting code, for $t\geq \ell\geq 1$, Sala et al. \cite{Sala-17-IT} showed that 
    \begin{gather}\label{eq:Sala}
        \nu(\mathcal{C},\mathcal{I}_t)\leq \frac{(q-1)^{t-\ell}\binom{2\ell}{\ell}}{(t-\ell)!}n^{t-\ell}+O(n^{t-\ell-1}).
    \end{gather}
    \item When $\mathcal{C}$ is an $(n,2,2;\mathcal{D}_1)$-reconstruction code, for $t\geq 2$, Chrisnata et al. \cite{Chrisnata-22-IT} established that $\nu(\mathcal{C},\mathcal{D}_t)\leq \frac{n^{t-1}}{(t-1)!} + O(n^{t-2})$, which follows by Equation (\ref{eq:t=1}) that $\rho(n,2,\frac{n^{t-1}}{(t-1)!} + O(n^{t-2});\mathcal{D}_t) \leq \log \log n + O(1)$. They also conjectured that this bound is tight, i.e., 
    \begin{gather}\label{eq:Chrisnata}
        \rho(n,2,\tfrac{n^{t-1}}{(t-1)!} + O(n^{t-2});\mathcal{D}_t)= \log \log n + O(1).
    \end{gather}
    \item When $\mathcal{C}$ is an $(n,2,2;\mathcal{I}_1)$-reconstruction code, for $t\geq 2$, Ye et al. \cite{Ye-23-IT} demonstrated that $\nu(\mathcal{C},\mathcal{I}_t)\leq \frac{n^{t-1}}{(t-1)!}+O(n^{t-2})$ and 
    \begin{gather}\label{eq:Ye}
        \rho(n,2,\tfrac{n^{t-1}}{(t-1)!} + O(n^{t-2});\mathcal{I}_t)= \log \log n + O(1).
    \end{gather}
    Recently, Abbasian et al. \cite{Abbasian-25-TMBMC} extended this result to non-binary alphabets for the special case where $t=2$.
\end{itemize}
It should be noted that all of the upper bounds for $ \nu(\mathcal{C}, \mathcal{B}) $ presented above are tight, as the corresponding references show the existence of codes that attain these bounds.

Recall that reconstruction codes are a generalization of error-correcting codes. It is well-known from \cite{Levenshtein-66} that there exists an equivalence between deletion-correcting codes and insertion-correcting codes. However, whether a similar connection exists between deletion reconstruction codes and insertion reconstruction codes has remained unknown in the literature. 
Our first contribution is to establish a direct connection. Specifically, by leveraging a relationship from the theory of longest common subsequences and shortest common supersequences, as established by Alon et al. \cite{Alon-24-IT}, we demonstrate that for any two distinct sequences, the intersection size of their $t$-deletion balls is no greater than the intersection size of their $t$-insertion balls. 
Consequently, we can establish the following results:
\begin{itemize}
  \item Any $(n, q, N; \mathcal{I}_t)$-reconstruction code is also an $(n, q, N; \mathcal{D}_t)$-reconstruction code.  
  \item We  have the inequality $\rho(n, q, N; \mathcal{D}_t) \leq \rho(n, q, N; \mathcal{I}_t)$. 
  \item Using Equation (\ref{eq:Sala}), we can generalize a key result from Pham et al. \cite{Pham-25-JCTA} (see Equation (\ref{eq:Pham})).
\end{itemize}
However, the other direction is typically untenable.
Our second contribution is to identify a significant distinction between reconstruction codes for deletions and insertions in the regime where $N=O(n^{t-1})$ and $t\geq 2$ is a constant.
\begin{itemize}
    \item For deletions, by considering the set of all sequences containing at most $\left\lfloor (q-1)(n-1)/q \right\rfloor+1$ runs, we establish that
        \begin{gather*}
          \rho\big(n,q,\tfrac{2(q-1)^{t-1}}{q^{t-1}(t-1)!}n^{t-1}+O(n^{t-2});\mathcal{D}_t\big)=O(1).
        \end{gather*}
         This disproves a conjecture posed by Chrisnata et al. \cite{Chrisnata-22-IT} (see Equation (\ref{eq:Chrisnata})).
    \item For insertions, by investigating the $(n,q,2;\mathcal{I}_1)$-reconstruction codes, we demonstrate that 
        \begin{gather*}
          \rho\big(n,q,\tfrac{(q-1)^{t-1}}{(t-1)!}n^{t-1}+O(n^{t-2});\mathcal{I}_t\big)=\log\log n + O(1).
        \end{gather*}
         This extends a key result from Ye et al. \cite{Ye-23-IT} (see Equation (\ref{eq:Ye})).
\end{itemize}
Finally, we explore the construction of $(n, q, N; \mathcal{I}_2)$-reconstruction codes, which are also $(n, q, N; \mathcal{D}_2)$-reconstruction codes, for $N \in \{2, 3, 4, 5\}$. According to existing literature, the first step involves characterizing pairs of $q$-ary sequences for which the intersection size of their two-insertion balls is at least $N$. Although such a characterization can be approached similarly to \cite{Sun-23-IT}, where binary sequences with fixed intersection sizes of their two-deletion balls are analyzed, constructing codes that exclude these pairs requires new ideas.
By imposing additional constraints on certain codes from \cite{Sun-25-arXiv, Sun-25-IT-C}, we construct $(n, q, N; \mathcal{I}_t)$-reconstruction codes with redundancies of $3 \log n + O(\log \log n)$, $3 \log n + O(1)$, $2 \log n + O(\log \log n)$, and $\log n + O(\log \log n)$ bits for $N=2,3,4,5$, respectively. 
When ignoring the $\log\log n$ items, these results align with the conclusions obtained for binary two-deletion reconstruction codes (see Equation (\ref{eq:t=2})). 
It is worth noting that for $N=1$ (i.e., classical two-insertion/deletion correcting codes), there remains a $\log n$ gap between existing constructions for $(n, 2, 1; \mathcal{D}_2)$-reconstruction codes and $(n, q, 1; \mathcal{I}_2)$-reconstruction codes with $q \geq 3$.

The rest of this paper is organized as follows.  
Section \ref{sec:notation} introduces relevant notations, formalizes the definition of reconstruction codes, and summarizes our main contributions. 
Section \ref{sec:connection} establishes the connection between $(n, q, N; \mathcal{D}_t)$-reconstruction codes and $(n, q, N; \mathcal{I}_t)$-reconstruction codes for any positive integers $N$ and $t$.  
Section \ref{sec:distinction} identifies a significant distinction between $(n, q, N; \mathcal{D}_t)$-reconstruction codes and $(n, q, N; \mathcal{I}_t)$-reconstruction codes in the regime where $N = O(n^{t-1})$ and $t \geq 2$ is a constant.  
Section \ref{sec:rec_code} presents the construction of $(n, q, N; \mathcal{I}_2)$-reconstruction codes and $(n, q, N; \mathcal{D}_2)$-reconstruction codes for $N \in \{2,3,4,5\}$.

\section{Notations and Problem Statement}\label{sec:notation}

\subsection{Basic Notations}
For integers $i, j$, let $[i, j]$ denote the set of integers $\{i, i+1, \dots, j\}$ if $i \leq j$, and the empty set $\varnothing$ otherwise.
For an integer $q\geq 2$, let $\Sigma_q$ be the $q$-ary alphabet $[0, q-1]$.  
In particular, when $q=2$, $\Sigma_2$ represents the binary alphabet $\{0, 1\}$.
Let $\Sigma_q^n$ (respectively, $\Sigma_q^{\geq n}$) be the set of all sequences of length exactly $n$ (respectively, at least $n$) over $\Sigma_q$.
Sequences are denoted by bold letters, while individual symbols within a sequence are denoted by plain letters.  
For any sequence $\bm{x} \in \Sigma_q^n$, we write $\bm{x} = x_1 x_2 \cdots x_n = (x_1, x_2, \dots, x_n)$, where $x_i$ represents the $i$-th entry of $\bm{x}$ for $i \in [1, n]$.
Given a set $\mathcal{S} = \{s_1, s_2, \dots, s_m\} \subseteq [1, n]$ with $s_1 < s_2 < \cdots < s_m$, we define $\bm{x}_{\mathcal{S}} = x_{s_1} x_{s_2} \cdots x_{s_m}$, which we term a \emph{subsequence} of $\bm{x}$. 
In the special case where $\mathcal{S}$ is an interval, $\bm{x}_{\mathcal{S}}$ is called a \emph{substring} of $\bm{x}$.
Let $|\bm{x}|$ denote the \emph{length} of the sequence $\bm{x}$ and let $|\mathcal{S}|$ denote the \emph{cardinality} of the set $\mathcal{S}$.  It is clear that $|\bm{x}_{\mathcal{S}}| = |\mathcal{S}|$.

For two sequences $\bm{x}, \bm{y} \in \Sigma_q^n$, their \emph{concatenation} is the sequence $x_1 \cdots x_n y_1 \cdots y_n$, denoted by $\bm{x}\bm{y}$ or $\bm{x} \circ \bm{y}$.
For two sets $\mathcal{S}, \mathcal{T} \subseteq \Sigma_q^n$, their \emph{concatenation} is the set $\{\bm{xy} : \bm{x} \in \mathcal{S}, \bm{y} \in \mathcal{T} \}$, denoted by $\mathcal{S} \circ \mathcal{T}$.  
In the special case where $\mathcal{S} = \{\bm{x}\}$ (respectively, $\mathcal{T} = \{\bm{y}\}$), we simplify the notation $\mathcal{S} \circ \mathcal{T}$ to $\bm{x} \circ \mathcal{T}$ (respectively, $\mathcal{S} \circ \bm{y}$).
For two symbols $a, b \in \Sigma_q$, let $\mathcal{S}^a = \{\bm{x} \in \mathcal{S} : x_1 = a\}$, $\mathcal{S}_b = \{\bm{x} \in \mathcal{S} : x_n = b\}$, and $\mathcal{S}_b^a = \{\bm{x} \in \mathcal{S} : x_1 = a, x_n = b\}$.

For any $t \in [1, n]$, we say that $\bm{x} \in \Sigma_q^n$ is $t$-\emph{periodic} if $x_i = x_{i+t}$ for $i \in [1, n-t]$, and that $\bm{x}$ is $t^{\leq}$-\emph{periodic} if there exists a $t' \in [1, t]$ such that $\bm{x}$ is $t'$-\emph{periodic}.
Note that any sequence of length at most $t$ is regarded as $t$-periodic.
A $2$-periodic sequence $\bm{x}$ is uniquely determined by its first two symbols (if it exists) and its length.
Therefore, we denote it by $\bm{\alpha}_n(x_1 x_2)$.
Let $\mathcal{R}(n, t, \ell)$ be the set of all sequences in $\Sigma_q^n$ such that every $t^{\leq}$-periodic substring has length at most $\ell$.

Throughout this paper, unless otherwise specified, for two distinct sequences $\bm{x}, \bm{y}\in \Sigma_q^n$, we will write $\bm{x} = \bm{u} \tilde{\bm{x}} \bm{v}$ and $\bm{y} = \bm{u} \tilde{\bm{y}} \bm{v}$ for some $\bm{u},\bm{v},\tilde{\bm{x}}, \tilde{\bm{y}} \in \Sigma_q^{\geq 0}$, where $\bm{u}$ and $\bm{v}$ denote the longest common prefix and suffix, respectively, of $\bm{x}$ and $\bm{y}$. 
Note that the sequences $\bm{u}$, $\bm{v}$, $\tilde{\bm{x}}$, and $\tilde{\bm{y}}$ are uniquely determined by $\bm{x}$ and $\bm{y}$. 
Moreover, the first (respectively, last) symbols of $\tilde{\bm{x}}$ and $\tilde{\bm{y}}$ are distinct.

\subsection{Reconstruction Codes}

We now formalize the definition of reconstruction codes.

\begin{definition}
Let $\mathcal{B}(\cdot)$ represent an error-ball function, and let $\mathcal{C}$ be a set of $q$-ary sequences of length $n$.  
The \emph{read coverage} of $\mathcal{C}$ with respect to $\mathcal{B}$, denoted by $\nu(\mathcal{C};\mathcal{B})$, is defined as
\begin{align*}
  \nu(\mathcal{C};\mathcal{B}) \triangleq \max_{\bm{x}\neq \bm{y}\in \mathcal{C}} \{|\mathcal{B}(\bm{x}) \cap \mathcal{B}(\bm{y})|\}.
\end{align*}
$\mathcal{C}$ is called an \emph{$(n,q,N;\mathcal{B})$-reconstruction code} if its read coverage satisfies $\nu(\mathcal{C};\mathcal{B}) < N$.
\end{definition}

Our primary focus is on characterizing the trade-off between code redundancy and read coverage.
To this end, we define the minimum redundancy of reconstruction codes.
Unless otherwise specified, all logarithms are taken to base $2$.

\begin{definition}
Let $\mathcal{C}$ be a $q$-ary code of length $n$. The \emph{redundancy} of $\mathcal{C}$, denoted by $r(\mathcal{C})$, is defined as
\begin{equation*}
    r(\mathcal{C}) \triangleq n \log q - \log |\mathcal{C}|.
\end{equation*}
Given positive integers $n, q, N$ and an error-ball function $\mathcal{B}(\cdot)$, the \emph{minimum redundancy} of an $(n,q,N;\mathcal{B})$-reconstruction code, denoted by $\rho(n,q,N;\mathcal{B})$, is given by
\begin{equation*}
  \rho(n,q,N;\mathcal{B}) = \min \{r(\mathcal{C}) : \mathcal{C} \subseteq \Sigma_q^n, \nu(\mathcal{C};\mathcal{B}) < N \}.
\end{equation*}
\end{definition}

We now introduce the notion of \emph{deletion ball} and \emph{insertion ball}, as we analyze channels that introduce deletions or insertions.

\begin{definition}
  For a given integer $t \geq 0$, let $\mathcal{D}_t(\cdot)$ denote the \emph{$t$-deletion ball} function, which maps a sequence to the set of all sequences obtainable by removing exactly $t$ symbols. Similarly, let $\mathcal{I}_t(\cdot)$ denote the \emph{$t$-insertion ball} function, which maps a $q$-ary sequence to the set of all sequences obtainable by inserting exactly $t$ symbols (from $\Sigma_q$).
\end{definition}

\begin{definition}
  Let $\bm{x}\in \Sigma_q^n$ and $\bm{y}\in \Sigma_q^m$, the \emph{Levenshtein distance} between them, denoted by $d_L(\bm{x},\bm{y})$, is the smallest number of insertion and deletion operations required to transform $\bm{x}$ into $\bm{y}$.
\end{definition}

For $\bm{x}, \bm{y}\in \Sigma_q^n$, it is well-known that 
\begin{equation}\label{eq:Levenshtein}
  \begin{aligned}
  d_L(\bm{x},\bm{y})
  &= \min\{2t: \mathcal{D}_t(\bm{x})\cap \mathcal{D}_t(\bm{y})\neq \varnothing\}\\
  &= \min\{2t: \mathcal{I}_t(\bm{x})\cap \mathcal{I}_t(\bm{y})\neq \varnothing\}.
  \end{aligned}
\end{equation}

The main objective addressed in this paper is:
\begin{problem}
    For integers $q, N, t$, determine or establish bounds and relationships for the minimum redundancies $\rho(n,q,N;\mathcal{D}_t)$ and $\rho(n,q,N;\mathcal{I}_t)$ as $n$ tends to infinity (i.e., for sufficiently large $n$).
\end{problem}

\subsection{Our Contributions}
We summarize our contributions as follows.
\begin{itemize}
  \item For arbitrary $n\geq t$, we show that $\rho(n,q,N;\mathcal{D}_t)\leq \rho(n,q,N;\mathcal{I}_t)$ (see Theorem \ref{thm:connection}).
  \item For $t\geq 2$ and for sufficiently large $n$, we prove that $\rho(n,q,\tfrac{2(q-1)^{t-1}}{q^{t-1}(t-1)!}n^{t-1}+O(n^{t-2});\mathcal{D}_t)\leq \log q$ and $\rho(n,q,\tfrac{(q-1)^{t-1}}{(t-1)!}n^{t-1}+O(n^{t-2});\mathcal{I}_t)=\log\log n + O(1)$ (see Theorem \ref{thm:distinction}).
  \item For $\mathcal{B}\in \{\mathcal{D}_2,\mathcal{I}_2\}$ and for sufficiently large $n$, we demonstrate that $\rho(n, q, 2; \mathcal{B})\leq 3\log n+O(\log\log n)$, $\rho(n, q, 3; \mathcal{B})\leq 3\log n+O(1)$, $\rho(n, q,4; \mathcal{B})\leq 2\log n+O(\log\log n)$, and $\rho(n, q,5; \mathcal{B})\leq \log n+O(\log\log n)$ (see Corollary \ref{cor:code}).
\end{itemize}

\section{The connection between reconstruction codes for deletions and insertions}\label{sec:connection}

In this section, we explore the connection between $(n, q, N; \mathcal{D}_t)$-reconstruction codes and $(n, q, N; \mathcal{I}_t)$-reconstruction codes. When $N=1$, the $(n, q, 1; \mathcal{D}_t)$-reconstruction codes and $(n, q, 1; \mathcal{I}_t)$-reconstruction codes essentially reduce to $t$-deletion correcting codes and $t$-insertion correcting codes, respectively. By \cite[Lemma 1]{Levenshtein-66}, we can derive the following connection between $(n, q, 1; \mathcal{D}_t)$-reconstruction codes and $(n, q, 1; \mathcal{I}_t)$-reconstruction codes.

\begin{lemma}
    A code $\mathcal{C} \subseteq \Sigma_q^n$ is an $(n, q, 1; \mathcal{D}_t)$-reconstruction code if and only if it is an $(n, q, 1; \mathcal{I}_t)$-reconstruction code.
\end{lemma}

It is natural to ask whether there exists a direct connection between $(n, q, N; \mathcal{D}_t)$-reconstruction codes and $(n, q, N; \mathcal{I}_t)$-reconstruction codes when $N\geq 2$.
An implicit connection can be derived from the theory of longest common subsequences and shortest common supersequences established by Alon et al. \cite{Alon-24-IT}, which can be stated as follows:

\begin{lemma}\cite[Proposition 2]{Alon-24-IT}
    Let $\bm{x}, \bm{y} \in \Sigma_q^n$ be such that $d_L(\bm{x}, \bm{y}) \geq 2t$, it holds that $|\mathcal{D}_t(\bm{x}) \cap \mathcal{D}_t(\bm{y})| \leq |\mathcal{I}_t(\bm{x}) \cap \mathcal{I}_t(\bm{y})|$.
    Consequently, for any integers $N, t \geq 1$, and any $(t-1)$-deletion/insertion correcting code $\mathcal{C}$, if $\mathcal{C}$ is an $(n, q, N; \mathcal{I}_t)$-reconstruction code, then $\mathcal{C}$ is also an $(n, q, N; \mathcal{D}_t)$-reconstruction code.
\end{lemma}

In the following, we extend this conclusion to arbitrary code $\mathcal{C} \subseteq \Sigma_q^n$.

\begin{theorem}\label{thm:connection}
    Let $\bm{x},\bm{y}\in \Sigma_q^n$, it holds that $|\mathcal{D}_t(\bm{x})\cap \mathcal{D}_t(\bm{y})|\leq |\mathcal{I}_t(\bm{x})\cap \mathcal{I}_t(\bm{y})|$.
    Consequently, for any integers $N,t\geq 1$, the following holds.
    \begin{itemize}
      \item If $\mathcal{C}$ is an $(n,q,N;\mathcal{I}_t)$-reconstruction code, then $\mathcal{C}$ is also an $(n,q,N;\mathcal{D}_t)$-reconstruction code;
      \item $\rho(n,q,N;\mathcal{D}_t)\leq \rho(n,q,N;\mathcal{I}_t)$.
    \end{itemize}
\end{theorem}

\begin{IEEEproof}
    For any $\bm{z}\in \mathcal{D}_t(\bm{x})\cap \mathcal{D}_t(\bm{y})$, we define the sets
    \begin{gather*}
        P(\bm{x},\bm{z})=\{p_1,p_2,\ldots,p_{n-t}\},\\
        P(\bm{y},\bm{z})=\{p_1',p_2',\ldots,p_{n-t}'\},
    \end{gather*}
    where $p_i$ (respectively, $p_i'$) is the smallest index such that $\bm{z}_{[1,i]}$ is a subsequence of $\bm{x}_{[1,p_i]}$ (respectively, $\bm{y}_{[1,p_i']}$), for $i\in [1,n-t]$.
    In other words, $p_i$ is the minimal position after $p_{i-1}$ such that $x_{p_i}=z_i$. 
    We then define the function $\varphi(\cdot)$ as follows:
    \begin{align*}
        \varphi(\bm{z})= \bm{x}_{[1,p_1-1]} \bm{y}_{[1,p_1']}\circ \bm{x}_{[p_1+1,p_2-1]} \bm{y}_{[p_1'+1,p_2']}\circ \cdots \circ \bm{x}_{[p_{n-t-1}+1,p_{n-t}-1]} \bm{y}_{[p_{n-t-1}'+1,p_{n-t}']} \circ \bm{x}_{[p_{n-t}+1,n]} \bm{y}_{[p_{n-t}'+1,n]}.
    \end{align*}
    It is clear that $\varphi(\bm{z})\in \mathcal{I}_t(\bm{x})\cap \mathcal{I}_t(\bm{y})$.
    It remains to show that the function $\varphi(\cdot)$ is an injection.
    
    For any $\tilde{\bm{z}}\in \mathcal{D}_t(\bm{x})\cap \mathcal{D}_t(\bm{y})$ with $\tilde{\bm{z}}\neq \bm{z}$, we define the sets $P(\bm{x},\tilde{\bm{z}})=\{\tilde{p}_1,\tilde{p}_2,\ldots,\tilde{p}_{n-t}\}$ and $P(\bm{y},\tilde{\bm{z}})=\{\tilde{p}_1',\tilde{p}_2',\ldots,\tilde{p}_{n-t}'\}$.
    Let $i$ be the smallest index at which $\tilde{\bm{z}}$ and $\bm{z}$ differ, then $p_i\neq \tilde{p}_i$, $p_i'\neq \tilde{p}_i'$, $p_{i-1}=\tilde{p}_{i-1}$, and $p_{i-1}'= \tilde{p}_{i-1}'$.
    Without loss of generality, we assume $p_i<\tilde{p}_i$.
    By definition, the substring $\varphi(\tilde{\bm{z}})_{[\tilde{p}_{i-1}+\tilde{p}_{i-1}'-i+2,\tilde{p}_i+\tilde{p}_i'-i-1]}= \bm{x}_{[\tilde{p}_{i-1}+1, \tilde{p}_{i}-1]} \bm{y}_{[\tilde{p}_{i-1}'+1,\tilde{p}_i'-1]}$ does not contain the symbol $\tilde{z}_i$, where $\tilde{z}_i=x_{\tilde{p}_i}=y_{\tilde{p}_i'}$.
    Observe that the substring $\varphi(\bm{z})_{[\tilde{p}_{i-1}+\tilde{p}_{i-1}'-i+2,\tilde{p}_i+\tilde{p}_i'-i-1]}$ contains either $x_{\tilde{p}_i}$ or $y_{\tilde{p}_i'}$. 
    It follows that $\varphi(\bm{z})\neq \varphi(\tilde{\bm{z}})$.
    This completes the proof. 
\end{IEEEproof}

Combining Theorem \ref{thm:connection} with Equations (\ref{eq:Sala}) and (\ref{eq:Levenshtein}), we can derive the following corollary.

\begin{corollary}\label{cor:JCTA}
  For any  $t\geq \ell\geq 0$, let $\bm{x},\bm{y}\in \Sigma_q^n$ be such that $d_L(\bm{x},\bm{y})\geq 2\ell$, it holds that 
  \begin{align*}
    |\mathcal{D}_t(\bm{x}) \cap \mathcal{D}_t(\bm{y})|\leq \frac{(q-1)^{t-\ell}\binom{2\ell}{\ell}}{(t-\ell)!}n^{t-\ell}+O(n^{t-\ell-1}).
  \end{align*}
\end{corollary}

\begin{remark}
     When $q=2$, Corollary \ref{cor:JCTA} confirms the conclusion of \cite[Theorem 4]{Pham-25-JCTA} (see Equation (\ref{eq:Pham})).
\end{remark}

\section{A significant distinction between reconstruction codes for deletions and insertions}\label{sec:distinction}

In the previous section, we established that $\rho(n, q, N; \mathcal{D}_t) \leq \rho(n, q, N; \mathcal{I}_t)$ for any $N$ and $t$. However, the reverse inequality generally does not hold. In the following, we will highlight a significant distinction between $\rho(n, q, N; \mathcal{D}_t)$ and $\rho(n, q, N; \mathcal{I}_t)$ in the regime where $N = O(n^{t-1})$ and $t \geq 2$ is a constant.

\begin{theorem}\label{thm:distinction}
    Let $q,t\geq 2$, for sufficiently large $n$, we have 
    \begin{gather*}
      \rho\big(n,q,\tfrac{2(q-1)^{t-1}}{q^{t-1}(t-1)!}n^{t-1}+O(n^{t-2});\mathcal{D}_t\big)\leq \log q, \\
      \rho\big(n,q,\tfrac{(q-1)^{t-1}}{(t-1)!}n^{t-1}+O(n^{t-2});\mathcal{I}_t\big)=\log\log n + O(1).
    \end{gather*}
\end{theorem}

\begin{remark}
    For deletions, Theorem \ref{thm:distinction} disproves a conjecture proposed by Chrisnata et al. \cite{Chrisnata-22-IT} (see Equation (\ref{eq:Chrisnata})).
    Moreover, for insertions, Theorem \ref{thm:distinction} generalizes a contribution of Ye et al. \cite{Ye-23-IT} (see Equation (\ref{eq:Ye})).
\end{remark}

The proof will be divided into two parts, corresponding to Theorems \ref{thm:del_case} and \ref{thm:ins_case}, which address deletions and insertions, respectively. Our strategies differ for each case. Specifically, for deletions, we consider the set of all sequences whose number of runs does not exceed $\frac{(q - 1)(n - 1)}{q} + 1$, where a \emph{run} in $\bm{x}$ is defined as a maximal substring of identical entries. Conversely, for insertions, we focus on the $(n, q, 2; \mathcal{I}_1)$-reconstruction codes.

Before proceeding with the proof, we list the following conclusions that will be used later.

\subsection{Useful Conclusions}

\subsubsection{The Deletion Ball}

It is well known that the size of a deletion ball depends on its center sequence and is strongly correlated with the number of runs within that sequence.
We denote by $r(\bm{x})$ the number of runs in $\bm{x}$.

\begin{lemma}\cite[Equation (11)]{Levenshtein-01-JCTA}\label{lem:del_run}
  For any $\bm{x}\in \Sigma_q^n$, it holds that $|\mathcal{D}_t(\bm{x})|\leq \binom{r(\bm{x})+t-1}{t}$.
\end{lemma}

For integers $n\geq t\geq 1$, since $\binom{n}{t}= \binom{n-1}{t}+ \binom{n-1}{t-1}$, we have 
\begin{equation}\label{eq:binom}
  \binom{n}{t}\geq \binom{n-1}{t} \text{ and } \binom{n}{t}\geq \binom{n-1}{t-1}.
\end{equation}

The following lemma, which appears in \cite[Proposition 16]{Chrisnata-22-IT} for the binary case, can be easily generalized to the non-binary case, as the proofs in \cite[Proposition 16]{Chrisnata-22-IT} do not depend on the size of the alphabet.

\begin{lemma}\cite[Proposition 16]{Chrisnata-22-IT}\label{lem:del_intersect}
  For any $\bm{x}= \bm{u} \tilde{\bm{x}} \bm{v}, \bm{y}= \bm{u} \tilde{\bm{y}} \bm{v} \in \Sigma_q^n$, it holds that
  \begin{align*}
    \mathcal{D}_t(\bm{x})\cap \mathcal{D}_t(\bm{y})
    = \bigcup_{t_1+t_2\leq t} \mathcal{D}_{t_1}(\bm{u}) \circ \big( \mathcal{D}_{t-t_1-t_2}(\tilde{\bm{x}})\cap \mathcal{D}_{t-t_1-t_2}(\tilde{\bm{y}}) \big) \circ \mathcal{D}_{t_2}(\bm{v}).
  \end{align*}
\end{lemma}

\subsubsection{The Insertion Ball}
Unlike deletion balls, the size of an insertion ball depends solely on its radius and the length of the center sequence, and is independent of the specific choice of that sequence.

\begin{lemma}\cite[Equations (24) and (25)]{Levenshtein-01-JCTA}\label{lem:ins_int'}
    Assume $t\geq 1$. For any $\bm{x}\in \Sigma_q^n$, it holds that $|\mathcal{I}_t(\bm{x})|=I_q(n,t)$, where
    \begin{equation}\label{eq:ins_def}
      \begin{gathered}
      I_q(n,t)= \sum_{i=0}^t \binom{n+t}{i}(q-1)^i=\frac{(q-1)^t}{t!} n^t+O(n^{t-1}),\\
      I_q(n,t)= I_q(n-1,t)+(q-1)I_q(n,t-1).
      \end{gathered}
    \end{equation}
\end{lemma}

For integers $t\geq k\geq \ell\geq 1$ and $q\geq 2$, we define
\begin{equation}\label{eq:ins_int_def}
  N_q^+(n,t,k,\ell)
  = \max_{\substack{\bm{x}\in \Sigma_q^{n+t-k},\bm{y}\in \Sigma_q^n\\ d_L(\bm{x},\bm{y})\geq t-k+2\ell}} \{|\mathcal{I}_k(\bm{x}) \cap \mathcal{I}_t(\bm{y})|\}.
\end{equation}
The following lemma determines the value of $N_q^+(n,t,k,\ell)$.
    
\begin{lemma}\cite[Theorem 5 and Lemma 6]{Sala-17-IT}\label{lem:ins_int}
  Assume $t\geq k\geq \ell\geq 1$ and $q\geq 2$, then
  \begin{align*}
    N_q^+(n,t,k,\ell)
    &= \sum_{j=\ell}^k \sum_{i=0}^{k-j} \binom{t-k+2j}{j} \binom{t+j-i}{t-k+2j} \binom{n+t}{i} (q-1)^i (-1)^{k+j-i}\\
    &=\frac{(q-1)^{k-\ell}\binom{t-k+2\ell}{\ell}}{(k-\ell)!}n^{k-\ell}+O(n^{k-\ell-1})
  \end{align*}
  and 
  \begin{align*}
      N_q^+(n,t,k,\ell)= N_q^+(n-1,t,k,\ell)+ (q-1)N_q^+(n,t-1,k-1,\ell).
  \end{align*}
\end{lemma}

The following lemma characterizes pairs of sequences whose single-insertion balls intersect in exactly one element.

\begin{lemma}\cite[Lemma 1 and Theorem 1]{Abbasian-25-TMBMC}\label{lem:char}
    Let $\bm{x}= \bm{u} \tilde{\bm{x}}\bm{v}, \bm{y}= \bm{u} \tilde{\bm{y}}\bm{v}\in \Sigma_q^n$ be such that $|\mathcal{I}_1(\bm{x}) \cap \mathcal{I}_1(\bm{y})|= 1$, one of the following holds:
    \begin{itemize}
      \item $\{\tilde{\bm{x}},\tilde{\bm{y}}\}=\{ac,cd\}$, where $a,c,d\in \Sigma_q$ are pairwise distinct (this case is only meaningful when $q\geq 3$);
      \item $\{\tilde{\bm{x}},\tilde{\bm{y}}\}=\{ac\bm{w}b,c\bm{w}bd\}$, where $a,b,c,d\in \Sigma_q$ and $\bm{w}\in \Sigma_q^{\geq 0}$ with $a\neq c,b\neq d$ and $ac\bm{w}\neq \bm{w}bd$.
    \end{itemize}
\end{lemma}

\subsection{The Deletion Part of Theorem \ref{thm:distinction}}\label{subsec:del}

\begin{theorem}\label{thm:del_case}
  Assume $t\geq 1$. Let $r=\left\lfloor (q-1)(n-1)/q \right\rfloor+1$ and $\mathcal{C}= \{\bm{x}\in \Sigma_q^n: r(\bm{x})\leq r\}$, then for any 
  \begin{align*}
      N
      &> 2\binom{r+t-3}{t-1}
      = \frac{2(q-1)^{t-1}n^{t-1}}{q^{t-1}(t-1)!}+O(n^{t-2}),
  \end{align*}
  $\mathcal{C}$ is an $(n,q,N;\mathcal{D}_t)$-reconstruction code with at most $\log q$ bits of redundancy. Consequently, we have $\rho(n,q,N;\mathcal{D}_t)\leq \log q$.
\end{theorem}

Our main idea is to establish an upper bound on the intersection size between two distinct deletion balls, expressed in terms of the number of runs in their respective center sequences.

\begin{lemma}\label{lem:del_int_run}
  Assume $t\geq 1$. For any two distinct sequences $\bm{x}, \bm{y}\in \Sigma_q^n$, let $\mathcal{S}= \mathcal{D}_t(\bm{x}) \cap \mathcal{D}_t(\bm{y})$, we have
  \begin{align*}
    |\mathcal{S}|\leq \binom{r(\bm{x})+t-3}{t-1}+ \binom{r(\bm{y})+t-3}{t-1}.
  \end{align*}
\end{lemma}

\begin{IEEEproof}
    We consider the following two cases.
    
    If $x_1\neq y_1$, let $i\geq 2$ and $j\geq 2$ be the smallest indices such that $x_i\neq x_1$ and $y_j=x_1$, respectively. Then, we have $r(\bm{x}_{[i,n]})\leq r(\bm{x})-1$ and $r(\bm{y}_{[j,n]})\leq r(\bm{y})-1$.
    Observe that
    \begin{gather*}
        \mathcal{S}^{x_1} 
        \subseteq \mathcal{D}_t(\bm{y})^{x_1}
        \subseteq \mathcal{D}_{t-j+1}(\bm{y}_{[j,n]}), \\
        \mathcal{S}\setminus \mathcal{S}^{x_1} 
        \subseteq \bigcup_{a' \in \Sigma_q\setminus x_1}\mathcal{D}_t(\bm{x})^{a'}
        \subseteq \mathcal{D}_{t-i+1}(\bm{x}_{[i,n]}).
    \end{gather*}
    We can compute
    \begin{equation*}
    \begin{aligned}
        |\mathcal{S}|
        &= |\mathcal{S}^{x_1}|+ |\mathcal{S}\setminus \mathcal{S}^{x_1} | \\
        &\leq |\mathcal{D}_{t-j+1}(\bm{y}_{[j,n]})|+ |\mathcal{D}_{t-i+1}(\bm{x}_{[i,n]})| \\
        &\stackrel{(\ast)}{\leq} \binom{r(\bm{y}_{[j,n]})+t-j}{t-j+1}+ \binom{r(\bm{x}_{[i,n]}) +t-i}{t-i+1} \\
        &\stackrel{(\star)}{\leq} \binom{r(\bm{y})+t-3}{t-1}+ \binom{r(\bm{x})+t-3}{t-1},
    \end{aligned}
    \end{equation*}
    where $(\ast)$ follows by Lemma \ref{lem:del_run} and $(\star)$ follows by Equation (\ref{eq:binom}).
    Thus, the conclusion holds for this case.
    
    If $x_1= y_1$, let $i\geq 2$ be the smallest index at which $\bm{x}$ and $\bm{y}$ differ, then $\bm{x}_{[i,n]}\neq \bm{y}_{[i,n]}$.
    Assume $r(\bm{x}_{[1,i-1]})= m\geq 1$, then $r(\bm{x}_{[i,n]})\leq r(\bm{x})-m+1$ and $r(\bm{y}_{[i,n]})\leq r(\bm{y})-m+1$. 
    By using the conclusion from the previous case, we get
    \begin{align*}
      |\mathcal{D}_t(\bm{x}_{[i,n]}) \cap \mathcal{D}_t(\bm{y}_{[i,n]})|
      &\leq \binom{r(\bm{y}_{[i,n]})+t-3}{t-1}+ \binom{r(\bm{x}_{[i,n]})+t-3}{t-1}\\
      &\leq \binom{r(\bm{y})-m+t-2}{t-1}+ \binom{r(\bm{x})-m+t-2}{t-1},
    \end{align*}
    where the last inequality follows by Equation (\ref{eq:binom}).
    Now, by Lemma \ref{lem:del_intersect}, we can compute
    \begin{equation*}
    \begin{aligned}
      |\mathcal{S}| 
      &\leq \sum_{k=0}^t |\mathcal{D}_k(\bm{x}_{[1,i-1]})| \cdot |\mathcal{D}_{t-k}(\bm{x}_{[i,n]}) \cap \mathcal{D}_{t-k}(\bm{y}_{[i,n]})| \\
      &\stackrel{(\ast)}{=} \sum_{k=0}^{t-1} |\mathcal{D}_k(\bm{x}_{[1,i-1]})| \cdot |\mathcal{D}_{t-k}(\bm{x}_{[i,n]}) \cap \mathcal{D}_{t-k}(\bm{y}_{[i,n]})| \\
      &\stackrel{(\star)}{\leq} \sum_{k=0}^{t-1} \binom{m+k-1}{k} \left[ \binom{r(\bm{y})-m+t-k-2}{t-k-1}+ \binom{r(\bm{x})-m+t-k-2}{t-k-1} \right]\\
      &\stackrel{(\diamond)}{=} \binom{r(\bm{y})+t-3}{t-1}+ \binom{r(\bm{x})+t-3}{t-1},
    \end{aligned}
    \end{equation*}
    where $(\ast)$ holds since $\bm{x}_{[i,n]}\neq \bm{y}_{[i,n]}$, $(\star)$ follows by Lemma \ref{lem:del_run}, and $(\diamond)$ is derived from the Cauchy-Vandermonde identity: $\binom{p_1+p_2}{s}=\sum_{k=0}^s \binom{p_1}{k} \binom{p_2}{s-k}$ (see \cite[Page 16, Exercise 1.9]{Jukna}).
    Thus, the conclusion also holds for this case.
    This completes the proof.
\end{IEEEproof}

We are now ready to prove Theorem \ref{thm:del_case}.
\begin{IEEEproof}[Proof of Theorem \ref{thm:del_case}]
    For any two distinct sequences $\bm{x},\bm{y}\in \mathcal{C}$, by Lemma \ref{lem:del_int_run}, we have $|\mathcal{D}_t(\bm{x})\cap \mathcal{D}_t(\bm{y})|<N$, which implies that $\mathcal{C}$ is an $(n,q,N;\mathcal{D}_t)$-reconstruction code.
    By \cite[Theorem 2]{Levenshtein-02-ISIT}, we have 
    \begin{align*}
        |\mathcal{C}|= q\sum_{i=0}^{r-1}\binom{n-1}{i}(q-1)^i.
    \end{align*}
    Next, we will demonstrate that $|\mathcal{C}|\geq q^{n-1}$ by examining the size of a $q$-ary covering code of length $n-1$ under the Hamming metric with a covering radius of $r-1$.
    A code $\mathcal{C}'\subseteq \Sigma_q^{n-1}$ is referred to as a covering code with a covering radius of $r-1$ if every sequence in $\Sigma_q^{n-1}$ is within a Hamming distance of at most $r-1$ from some codeword in $\mathcal{C}'$, where the Hamming distance between two sequences is defined as the number of positions at which the corresponding symbols differ.
    By \cite[Table 1]{Lenz-21-IT}, we have
    \begin{align*}
        |\mathcal{C}'|\geq \frac{q^{n-1}}{\sum_{i=0}^{r-1}\binom{n-1}{i}(q-1)^i}.
    \end{align*}
    Observe that for any $\bm{z}\in \Sigma_q^{n-1}$, by the pigeonhole principle, $\bm{z}$ contains at least $\lceil \frac{n-1}{q}\rceil$ occurrences of some symbol $a\in \Sigma_q$.
    In other words, the Hamming distance between $\bm{z}$ and $a^{n-1}$ is at most $n-1-\lceil \frac{n-1}{q}\rceil= r-1$. 
    This implies that the code $\{a^{n-1}: a\in \Sigma_q\}$ is a covering code with a covering radius of $r-1$.
    Therefore, we get
    \begin{align*}
        q\geq \frac{q^{n-1}}{\sum_{i=0}^{r-1}\binom{n-1}{i}(q-1)^i},
    \end{align*}
    which enables us to obtain $|\mathcal{C}|= q\sum_{i=0}^{r-1}\binom{n-1}{i}(q-1)^i\geq q^{n-1}$.
    Then the conclusion follows.
\end{IEEEproof}

\subsection{The Insertion Part of Theorem \ref{thm:distinction}}\label{subsec:ins}

Let
\begin{align*}
    \Delta_q(n,t)
    &\triangleq N_q^+(n-1,t-1,t-1,1)+ 2(q-2) N_q^+(n-1,t-1,t-2,1)\\
    &~~~~+ (q-2)^2 N_q^+(n,t-2,t-2,1)- (q-1)^2 I_q(n+1,t-3)\\
    &= \frac{2(q-1)^{t-2}}{(t-2)!}n^{t-2}+ O(n^{t-3}).
\end{align*}
By Lemmas \ref{lem:ins_int'} and \ref{lem:ins_int}, we get 
\begin{gather}\label{eq:delta}
  \Delta_q(n,t)= \Delta_q(n-1,t)+(q-1)\Delta_q(n,t-1).
\end{gather}
    
\begin{theorem}\label{thm:ins_case}
    Assume $t\geq 2$, let $n$ be such that $I_q(n+1,t-1)>\max\{\Delta_q(n,t)- N_q^+(n+1,t-1,t-1,1),N_q^+(n, t, t, 2)\}$ and let $N=\tfrac{(q-1)^{t-1}}{(t-1)!}n^{t-1}+O(n^{t-2})\in [I_q(n+1,t-1)+\Delta_q(n,t)+1, 2I_q(n+1,t-1)-N_q^+(n+1,t-1,t-1,1)]$, then 
    $\mathcal{C}$ is an $(n,q,2;\mathcal{I}_1)$-reconstruction code if and only if it is an $(n, q, N; \mathcal{I}_t)$-reconstruction code.
    This implies that $\rho(n,q,N;\mathcal{I}_t)=\rho(n,q,2;\mathcal{I}_1)=\log\log n+O(1)$.
\end{theorem}

\begin{proof}
  If $\mathcal{C}$ is an $(n, q, 2; \mathcal{I}_1)$-reconstruction code, for any two distinct sequences $\bm{x}, \bm{y} \in \mathcal{C}$, it holds that $|\mathcal{I}_1(\bm{x}) \cap \mathcal{I}_1(\bm{y})| \leq 1$.
  If $|\mathcal{I}_1(\bm{x}) \cap \mathcal{I}_1(\bm{y})| = 0$, then $d_L(\bm{x}, \bm{y}) \geq 4$. By Equation (\ref{eq:ins_int_def}), we can deduce that $|\mathcal{I}_t(\bm{x}) \cap \mathcal{I}_t(\bm{y})| \leq N_q^+(n, t, t, 2)< I_q(n+1, t-1)$.
  If $|\mathcal{I}_1(\bm{x}) \cap \mathcal{I}_1(\bm{y})| = 1$, then there exists some $\bm{z} \in \Sigma_q^{n+1}$ such that $\{\bm{z}\} = \mathcal{I}_1(\bm{x}) \cap \mathcal{I}_1(\bm{y})$. Observe that $\mathcal{I}_{t-1}(\bm{z}) \subseteq \mathcal{I}_t(\bm{x}) \cap \mathcal{I}_t(\bm{y})$ and $|\mathcal{I}_{t-1}(\bm{z})| = I_q(n+1, t-1)$. We will show in Lemma \ref{lem:ins_case} that $|\mathcal{I}_t(\bm{x}) \cap \mathcal{I}_t(\bm{y}) \setminus \mathcal{I}_{t-1}(\bm{z})| \leq \Delta_q(n, t)$, which implies $|\mathcal{I}_t(\bm{x}) \cap \mathcal{I}_t(\bm{y})| \leq I_q(n+1, t-1) + \Delta_q(n, t)<N$. Then $\mathcal{C}$ is an $(n, q, N; \mathcal{I}_t)$-reconstruction code.
  
  For the other direction, let $\mathcal{C}$ be an $(n, q, N; \mathcal{I}_t)$-reconstruction code, for any two distinct sequences $\bm{x},\bm{y}\in \mathcal{C}$, we can conclude that $|\mathcal{I}_1(\bm{x}) \cap \mathcal{I}_1(\bm{y})|\leq 1$.
  Otherwise, there exist two distinct sequences $\bm{z},\bm{z}'\in \mathcal{I}_1(\bm{x}) \cap \mathcal{I}_1(\bm{y})$ and we can compute 
  \begin{align*}
    |\mathcal{I}_t(\bm{x}) \cap \mathcal{I}_t(\bm{y})|
    &\geq |\mathcal{I}_{t-1}(\bm{z}) \cup \mathcal{I}_{t-1}(\bm{z}')|\\
    &= |\mathcal{I}_{t-1}(\bm{z})|+ |\mathcal{I}_{t-1}(\bm{z}')|- |\mathcal{I}_{t-1}(\bm{z}) \cap \mathcal{I}_{t-1}(\bm{z}')|\\
    &\geq 2I_q(n+1,t-1)-N_q^+(n+1,t-1,t-1,1)\\
    &\geq N,
  \end{align*}
  which leads to a contradiction.
  Thus, $\mathcal{C}$ is an $(n, q, 2; \mathcal{I}_1)$-reconstruction code.
  This completes the proof.
\end{proof}

\begin{lemma}\label{lem:ins_case}
    Assume $n,t\geq 2$. Let $\bm{x}= \bm{u}\tilde{\bm{x}}\bm{v}, \bm{y}=\bm{u}\tilde{\bm{y}}\bm{v} \in\Sigma_q^n$ be such that $\mathcal{I}_1(\bm{x}) \cap \mathcal{I}_1(\bm{y})= \{\bm{z}\}$, then 
    \begin{align*}
      |\mathcal{I}_t(\bm{x}) \cap \mathcal{I}_t(\bm{y}) \setminus \mathcal{I}_{t-1}(\bm{z})|
      &\leq\Delta_q(n,t)=O(n^{t-2}).
    \end{align*}
\end{lemma}

\begin{IEEEproof}
    Let $\mathcal{S}= \mathcal{I}_t(\bm{x}) \cap \mathcal{I}_t(\bm{y})$, we will prove $|\mathcal{S}\setminus \mathcal{I}_{t-1}(\bm{z})|\leq\Delta_q(n,t)$ by induction on $n+t$.
    
    For the base case where $n+t=4$, we have $n=t=2$ (we will consider the general case where $t\geq 2$, as it will be used later).
    In this case, by Lemma \ref{lem:char}, we can express $\bm{x}=ac$ and $\bm{y}=cd$, where $a,c,d\in \Sigma_q$ are pairwise distinct.
    Then, we have $\bm{z}= acd$.
    Observe that $\mathcal{S}=  \mathcal{S}^a \cup \mathcal{S}_d \cup (\cup_{a'\neq a, d'\neq d} S_{d'}^{a'})$, where
    \begin{gather*}
        \mathcal{S}^{a}
        \subseteq \mathcal{I}_t(\bm{y})^a
        =a \circ \mathcal{I}_{t-1}(cd)\subseteq \mathcal{I}_{t-1}(acd),\\
        \mathcal{S}_{d}
        \subseteq \mathcal{I}_t(\bm{x})_d
        =\mathcal{I}_{t-1}(ac) \circ d
        \subseteq \mathcal{I}_{t-1}(acd),\\
        \mathcal{S}_{d'}^{a'}=
        \begin{cases}
            a' \circ (\mathcal{I}_{t-1}(a)\cap \mathcal{I}_{t-1}(d)) \circ d', &\mbox{if } a'=c, d'=c; \\
            a'\circ (\mathcal{I}_{t-2}(ac)\cap \mathcal{I}_{t-1}(d)) \circ d', &\mbox{if } a'=c, d'\not\in \{c,d\}; \\ 
            a' \circ (\mathcal{I}_{t-1}(a)\cap \mathcal{I}_{t-2}(cd)) \circ d', &\mbox{if } a'\not\in \{a,c\}, d'=c; \\ 
            a' \circ (\mathcal{I}_{t-2}(ac)\cap \mathcal{I}_{t-2}(cd)) \circ d', &\mbox{if } a'\not\in \{a,c\}, d'\not\in \{c,d\}.
        \end{cases}
    \end{gather*}
    Moreover, for any $a'\neq a$ and $d'\neq d$, we have 
    \begin{gather*}
    \mathcal{S}_{d'}^{a'}\cap \mathcal{I}_{t-1}(acd)= a' \circ \mathcal{I}_{t-3}(acd) \circ d'.
    \end{gather*}
    It can be easily checked that $d_L(a,d)=2$, $d_L(ac,d)=3$, $d_L(a,cd)=3$, and $d_L(ac,cd)=2$.
    By Equation (\ref{eq:ins_int_def}), we obtain 
    \begin{gather*}
      |\mathcal{I}_{t-1}(a)\cap \mathcal{I}_{t-1}(d)|\leq N_q^+(1,t-1,t-1,1),\\
      |\mathcal{I}_{t-2}(ac)\cap \mathcal{I}_{t-1}(d)|\leq N_q^+(1,t-1,t-2,1),\\
      |\mathcal{I}_{t-1}(a)\cap \mathcal{I}_{t-2}(cd)|\leq N_q^+(1,t-1,t-2,1),\\
      |\mathcal{I}_{t-2}(ac)\cap \mathcal{I}_{t-2}(cd)|\leq N_q^+(2,t-2,t-2,1).
    \end{gather*}
    Then we can compute
    \begin{align*}
      \big|\mathcal{S}\setminus \mathcal{I}_{t-1}(acd) \big|
      &= \sum_{a'\in \Sigma_q\setminus \{a\}} \sum_{d'\in \Sigma_q\setminus \{d\}} \big|\mathcal{S}_{d'}^{a'}\setminus (a' \circ \mathcal{I}_{t-3}(acd) \circ d')\big|\\
      &= \sum_{a'\in \Sigma_q\setminus \{a\}} \sum_{d'\in \Sigma_q\setminus \{d\}}  \big|\mathcal{S}_{d'}^{a'}\big|- (q-1)^2 \big|\mathcal{I}_{t-3}(acd)\big|\\
      &\leq N_q^+(1,t-1,t-1,1)+ 2(q-2) N_q^+(1,t-1,t-2,1)\\
      &~~~~+ (q-2)^2 N_q^+(2,t-2,t-2,1)- (q-1)^2 I_q(3,t-3)\\
      &= \Delta_q(2,t).
    \end{align*}
    Consequently, the conclusion is valid for the base case.
    
    Now we assume the conclusion is valid for $n+t<n'$, where $n'\geq 5$, and examine the scenario where $n+t=n'$. We consider two cases.
    \begin{itemize}
      \item When $|\bm{u}|=|\bm{v}|=0$, since the case where $n=2$ has been discussed previously, it suffices to consider the case where $n\geq 3$.
      By Lemma \ref{lem:char}, we can express $\bm{x}=ac\bm{w}b$ and $\bm{y}=c\bm{w}bd$, where $a,b,c,d\in \Sigma_q$ and $\bm{w}\in \Sigma_q^{\geq 0}$ with $a\neq c,b\neq d$, and $ac\bm{w}\neq \bm{w}bd$. Then, we have $\bm{z}=ac\bm{w}bd$. 
        Observe that $\mathcal{S}=  \mathcal{S}^a \cup \mathcal{S}_d \cup (\cup_{a'\neq a, d'\neq d} S_{d'}^{a'})$, where
        \begin{gather*}
            \mathcal{S}^{a}
            \subseteq \mathcal{I}_t(\bm{y})^a
            =a \circ \mathcal{I}_{t-1}(c\bm{w}bd)
            \subseteq \mathcal{I}_{t-1}(ac\bm{w}bd),\\
          \mathcal{S}_{d}
          \subseteq \mathcal{I}_t(\bm{x})_d
          =\mathcal{I}_{t-1}(ac\bm{w}b) \circ d
          \subseteq \mathcal{I}_{t-1}(ac\bm{w}bd),\\
            \mathcal{S}_{d'}^{a'}=
            \begin{cases}
                a' \circ (\mathcal{I}_{t-1}(ac\bm{w})\cap \mathcal{I}_{t-1}(\bm{w}bd)) \circ d', &\mbox{if } a'=c, d'=b; \\
                a' \circ (\mathcal{I}_{t-2}(ac\bm{w}b)\cap \mathcal{I}_{t-1}(\bm{w}bd)) \circ d', &\mbox{if } a'=c, d'\not\in \{b,d\};\\
                a' \circ (\mathcal{I}_{t-1}(ac\bm{w})\cap \mathcal{I}_{t-2}(c\bm{w}bd)) \circ d', &\mbox{if } a'\not\in \{a,c\}, d'=b; \\ 
                a' \circ (\mathcal{I}_{t-2}(ac\bm{w}b)\cap \mathcal{I}_{t-2}(c\bm{w}bd)) \circ d', &\mbox{if } a'\not\in \{a,c\}, d'\not\in \{b,d\}.
            \end{cases}
        \end{gather*}
        Moreover, for any $a'\neq a$ and $d'\neq d$, we have 
        \begin{align*}
        \mathcal{S}_{d'}^{a'}\cap \mathcal{I}_{t-1}(ac\bm{w}bd)= a' \circ \mathcal{I}_{t-3}(ac\bm{w}bd) \circ d'.
        \end{align*} 
        It can be easily checked that $d_L(ac\bm{w},\bm{w}bd)\geq 2$, $d_L(ac\bm{w}b,\bm{w}bd)=3$, $d_L(ac\bm{w},c\bm{w}bd)=3$, and $d_L(ac\bm{w}b,c\bm{w}bd)=2$, then by Equation (\ref{eq:ins_int_def}), we obtain 
        \begin{gather*}
          \big|\mathcal{I}_{t-1}(ac\bm{w})\cap \mathcal{I}_{t-1}(\bm{w}bd)\big|\leq N_q^+(n-1,t-1,t-1,1),\\
          \big|\mathcal{I}_{t-2}(ac\bm{w}b)\cap \mathcal{I}_{t-1}(\bm{w}bd)\big|\leq N_q^+(n-1,t-1,t-2,1),\\
          \big|\mathcal{I}_{t-1}(ac\bm{w})\cap \mathcal{I}_{t-2}(c\bm{w}bd)\big|\leq N_q^+(n-1,t-1,t-2,1),\\
          \big|\mathcal{I}_{t-2}(ac\bm{w}b)\cap \mathcal{I}_{t-2}(c\bm{w}bd)\big|\leq N_q^+(n,t-2,t-2,1).
        \end{gather*}
        Thus, we can compute
        \begin{equation}\label{eq:t1}
        \begin{aligned}
          \big|\mathcal{S}\setminus \mathcal{I}_{t-1}(ac\bm{w}bd)\big|
          &= \sum_{a'\in \Sigma_q\setminus \{a\}} \sum_{d'\in \Sigma_q\setminus \{d\}}  \big|\mathcal{S}_{d'}^{a'}\setminus (a' \circ \mathcal{I}_{t-3}(ac\bm{w}bd) \circ d')\big|\\
          &= \sum_{a'\in \Sigma_q\setminus \{a\}} \sum_{b\in \Sigma_q\setminus \{d\}}  \big|\mathcal{S}_{d'}^{a'}\big|- (q-1)^2 \big|\mathcal{I}_{t-3}(ac\bm{w}bd)\big|\\
          &\leq N_q^+(n-1,t-1,t-1,1)+ 2(q-2) N_q^+(n-1,t-1,t-2,1)\\
          &~~~~+ (q-2)^2 N_q^+(n,t-2,t-2,1)- (q-1)^2 I_q(n+1,t-3)\\
          &= \Delta_q(n,t).
        \end{aligned}
        \end{equation}
      \item When $|\bm{u}|+|\bm{v}|\geq 1$, without loss of generality, we assume $|\bm{u}|\geq 1$. 
      In this case, we have $z_1=u_1$ and $\mathcal{I}_1(\bm{x}_{[2,n]}) \cap \mathcal{I}_1(\bm{y}_{[2,n]})=\{\bm{z}_{[2,n+1]}\}$. Observe that
            \begin{equation*}
                \big(\mathcal{S}\setminus \mathcal{I}_{t-1}(\bm{z})\big)^{a'}=
                \begin{cases}
                    a' \circ \big(\mathcal{I}_t(\bm{x}_{[2,n]}) \cap \mathcal{I}_t(\bm{y}_{[2,n]})\setminus \mathcal{I}_{t-1}(\bm{z}_{[2,n+1]})\big), &\mbox{if } a'=u_1; \\
                    a'\circ \big(\mathcal{I}_{t-1}(\bm{x}) \cap \mathcal{I}_{t-1}(\bm{y})\setminus \mathcal{I}_{t-2}(\bm{z}) \big), &\mbox{if } a'\neq u_1.
                \end{cases}
            \end{equation*}
        Then we can compute
        \begin{align*}
            |\mathcal{S}\setminus \mathcal{I}_{t-1}(\bm{z})|
            &= \big|\big(\mathcal{S}\setminus \mathcal{I}_{t-1}(\bm{z})\big)^{u_1}\big|+ \sum_{a'\in \Sigma_q\setminus \{u_1\}} \big| \big(\mathcal{S}\setminus \mathcal{I}_{t-1}(\bm{z})\big)^{a'} \big| \\
            &= \big|\mathcal{I}_t(\bm{x}_{[2,n]}) \cap \mathcal{I}_t(\bm{y}_{[2,n]})\setminus \mathcal{I}_{t-1}(\bm{z}_{[2,n+1]}) \big|+ (q-1) \cdot \big|\mathcal{I}_{t-1}(\bm{x}) \cap \mathcal{I}_{t-1}(\bm{y})\setminus \mathcal{I}_{t-2}(\bm{z}) \big| \\
            &\stackrel{(\ast)}{\leq} \Delta_q(n-1,t)+ (q-1)\Delta_q(n,t-1) \\
            &\stackrel{(\star)}{=}\Delta_q(n,t),
        \end{align*}
        where $(\ast)$ follows by the induction hypothesis and $(\star)$ follows by Equation (\ref{eq:delta}).
    \end{itemize}
    Consequently, the conclusion is also valid for the case where $n+t=n'$.
    This completes the proof.
\end{IEEEproof}

\begin{remark}
  In the proof of Lemma \ref{lem:ins_case}, when considering the sequences 
  $\bm{x}=ac\bm{w}b$ and $\bm{y}=c\bm{w}bd$, where $a,b,c,d\in \Sigma_q$ and $\bm{w}\in \Sigma_q^{\geq 0}$ with $a\neq c,b\neq d$, and $ac\bm{w}\neq \bm{w}bd$, we can establish a tighter bound on the condition that $\mathcal{I}_1(ac\bm{w})\cap \mathcal{I}_1(\bm{w}bd)= \varnothing$.
   In this case, we have $d_L(ac\bm{w},\bm{w}bd)\geq 4$, which implies that
    \begin{gather*}
        |\mathcal{I}_{t-1}(ac\bm{w})\cap \mathcal{I}_{t-1}(\bm{w}bd)|\leq N_q^+(n-1,t-1,t-1,2).
    \end{gather*}
   Then we can establish the following tighter bound for Equation (\ref{eq:t1}):
    \begin{align*}
        |\mathcal{S}\setminus \mathcal{I}_{t-1}(ac\bm{w}bd)|
            &\leq N_q^+(n-1,t-1,t-1,2)+ 2(q-2) N_q^+(n-1,t-1,t-2,1)\\
            &~~~~+ (q-2)^2 N_q^+(n,t-2,t-2,1)- (q-1)^2 I_q(n+1,t-3)\\
            &\triangleq \Delta_q'(n,t)= O(n^{t-3}).
    \end{align*}
    Moreover, by following the remaining discussion in Lemma \ref{lem:ins_case}, we can obtain:
    \begin{lemma}\label{lem:ins_case'}
        Assume $t\geq 2$. Let $\bm{x}= \bm{u}ac\bm{w}b\bm{v}, \bm{y}=\bm{u}c\bm{w}bd\bm{v} \in\Sigma_q^n$, where $a,b,c,d\in \Sigma_q$ and $\bm{u},\bm{v},\bm{w}\in \Sigma_q^{\geq 0}$ with $a\neq c$ and $b\neq d$, be such that  $\mathcal{I}_1(\bm{x})\cap \mathcal{I}_1(\bm{y})= \{\bm{z}\}$ and $\mathcal{I}_1(ac\bm{w})\cap \mathcal{I}_1(\bm{w}bd)= \varnothing$, then
        \begin{gather*}
          |\mathcal{I}_t(\bm{x})\cap \mathcal{I}_t(\bm{y})\setminus \mathcal{I}_{t-1}(\bm{z})|\leq \Delta_q'(n,t)=O(n^{t-3}).
        \end{gather*}
    \end{lemma}
\end{remark}

\begin{remark}
  When $q=2$ and $t\geq 3$, the upper bound on $|\mathcal{I}_t(\bm{x})\cap \mathcal{I}_t(\bm{y})|$ provided in Lemma \ref{lem:ins_case} is tighter than the bound given in \cite[Lemma VI.1]{Ye-23-IT}.
\end{remark}

\section{Constructions of Reconstruction codes for two deletions/insertions}\label{sec:rec_code}

In this section, we focus on constructing $(n, q, N; \mathcal{B})$-reconstruction codes for $\mathcal{B} \in \{\mathcal{D}_2, \mathcal{I}_2\}$ and $N \in \{2, 3, 4, 5\}$.  
By Theorem \ref{thm:connection}, it suffices to consider the case where $\mathcal{B} = \mathcal{I}_2$.  
Our main contribution can be summarized as follows:

\begin{theorem}\label{thm:code}
When $N$ is set to $2, 3, 4,$ and $5$, there exist $(n, q, N; \mathcal{I}_2)$-reconstruction codes with redundancies of $3\log n + O(\log \log n)$, $3\log n + O(1)$, $2\log n + O(\log \log n)$, and $\log n + O(\log \log n)$, respectively.
\end{theorem}

The proof will be divided into four parts, corresponding to Theorems \ref{thm:N=5}, \ref{thm:N=4}, \ref{thm:N=3}, and \ref{thm:N=2}, which address the cases where $N=5,4,3,2$, respectively.
By Theorem \ref{thm:connection}, the following statement holds.

\begin{corollary}\label{cor:code}
  Let $\mathcal{B}\in \{\mathcal{D}_2,\mathcal{I}_2\}$. When $N$ is set to $2$, $3$, $4$, and $5$, there exist $(n,q,N;\mathcal{B})$-reconstruction codes with redundancies of $3\log n+O(\log\log n)$, $3\log n+O(1)$, $2\log n+O(\log\log n)$, and $\log n+O(\log\log n)$, respectively. Then, we have
    \begin{equation*}
      \rho(n,q,N;\mathcal{B})\leq 
      \begin{cases}
        3\log n+O(\log\log n), & \mbox{if } N=2; \\
        3\log n+O(1), & \mbox{if } N=3;\\
        2\log n+O(\log \log n), & \mbox{if } N=4; \\
        \log n+O(\log\log n), & \mbox{if } N\geq 5.
      \end{cases}
    \end{equation*}
\end{corollary}

\begin{remark}
  It is noteworthy that, for $N=1$ (i.e., classical insertion/deletion-correcting codes), there exists a $\log n$ gap in existing constructions between $(n, 2, N; \mathcal{B})$-reconstruction codes and $(n, q, N; \mathcal{B})$-reconstruction codes, where $\mathcal{B}\in \{\mathcal{D}_2,\mathcal{I}_2\}$ and $q \geq 3$. In contrast, for $N \in \{2, 3, 4, 5\}$, as shown in Corollary \ref{cor:code}, we construct non-binary reconstruction codes that are at most $O(\log\log n)$ larger than their binary counterparts (see Equation (\ref{eq:t=2})).
\end{remark}

\subsection{Characterizations}

To construct $(n,q,N;\mathcal{I}_2)$-reconstruction codes, the first step is to characterize a pair of sequences whose two-insertion balls intersect at least $N$ elements.
We observe that the conclusions in \cite[Section III]{Sun-23-IT}, which characterize binary sequences with fixed intersection sizes of their two-deletion balls, can be generalized to insertion channels, although some minor modifications are required.
 Below, we explain how to extend these conclusions.

We consider two sequences $\bm{x}=\bm{u}\tilde{\bm{x}}\bm{v},\bm{y}=\bm{u}\tilde{\bm{y}}\bm{v}\in \Sigma_q^n$ with $d_L(\bm{x},\bm{y})\geq 4$, i.e., $d_L(\tilde{\bm{x}},\tilde{\bm{y}})\geq 4$.
Let $m\triangleq |\tilde{\bm{x}}|$, we define
\begin{equation}\label{eq:A}
\begin{gathered}
    \tilde{\bm{x}}(\ell,k)\triangleq \{\bm{z}\in \mathcal{I}_2(\tilde{\bm{x}}): \bm{z}_{[1,m+2]\setminus \{\ell,k\}}=\tilde{\bm{x}} \},\\
    \mathcal{A}_1 \triangleq \{ \bm{z}:
    \bm{z}\in \tilde{\bm{x}}(\ell,k)\cap \tilde{\bm{y}}(1,m+2),~1<\ell< k<m+2\}, \\
    \mathcal{A}_2 \triangleq \{ \bm{z}:
    \bm{z} \in \tilde{\bm{x}}(\ell,m+2)\cap \tilde{\bm{y}}(1,k),~1<\ell< k<m+2\}, \\
    \mathcal{A}_3 \triangleq \{ \bm{z}:
    \bm{z} \in \tilde{\bm{x}}(k,m+2)\cap \tilde{\bm{y}}(1,\ell),~1<\ell< k<m+2\}, \\
    \mathcal{A}_4 \triangleq \{ \bm{z}:
    \bm{z}\in \tilde{\bm{x}}(1,m+2)\cap \tilde{\bm{y}}(\ell,k),~1<\ell< k<m+2\}, \\
    \mathcal{A}_5 \triangleq \{ \bm{z}:
    \bm{z} \in \tilde{\bm{x}}(1,k)\cap \tilde{\bm{y}}(\ell,m+2),~1<\ell< k<m+2\}, \\
    \mathcal{A}_6 \triangleq \{ \bm{z}:
    \bm{z} \in \tilde{\bm{x}}(1,\ell)\cap \tilde{\bm{y}}(k,m+2),~1<\ell< k<m+2\}.
\end{gathered}
\end{equation}

\begin{lemma}\label{lem:classify}
    For two distinct sequences $\tilde{\bm{x}}= a \bm{r} b, \tilde{\bm{y}}= a' \bm{s} b'\in \Sigma_q^m$, where $m\geq 2$ and $a, a', b, b'\in \Sigma_q$ with $a\neq a'$ and $b \neq b'$, if $d_L(\tilde{\bm{x}},\tilde{\bm{y}}) \geq 4$, let $\mathcal{S}=\mathcal{I}_2(\tilde{\bm{x}}) \cap \mathcal{I}_2(\tilde{\bm{y}})$, then $|\mathcal{S}|= \cup_{i=1}^6 \mathcal{A}_i$ and $|\mathcal{A}_i| \leq 1$ for any $i\in [1,6]$.
\end{lemma}

Before proceeding with the proof, we require the following characterization of the intersection of single-insertion balls, which can be readily verified based on \cite[Definition 8 and Proposition 9]{Cai-22-IT}.

\begin{lemma}\label{lem:position}
    Let $\bm{x},\bm{y}\in \Sigma_q^n$ be distinct sequences, then $|\mathcal{I}_1(\bm{x})\cap \mathcal{I}_1(\bm{y})|\leq 2$.
    Moreover, if $\mathcal{I}_1(\bm{x})\cap \mathcal{I}_1(\bm{y})=\{\bm{z},\bm{z}'\}$, let $\ell$ and $k$ be the smallest and largest indices at which $\bm{x}$ and $\bm{y}$ differ, respectively, the following holds:
    \begin{itemize}
      \item $\bm{z}=\bm{x}_{[1,\ell-1]} y_{\ell} \bm{x}_{[\ell,n]}= \bm{y}_{[1,k]} x_k \bm{y}_{[k+1,n]}$;
      \item $\bm{z}'=\bm{x}_{[1,k]}y_k \bm{x}_{[k+1,n]}= \bm{y}_{[1,\ell-1]}x_{\ell}\bm{y}_{[\ell,n]}$.
    \end{itemize}
\end{lemma}

\begin{IEEEproof}[Proof of Lemma \ref{lem:classify}]
Let $\mathcal{S}=\mathcal{I}_{2}(\tilde{\bm{x}}) \cap \mathcal{I}_{2}(\tilde{\bm{y}})$.
We first show that $\mathcal{S}^{a''}=\varnothing$ for $a''\not\in \{a,a'\}$.
Assume there exists some $a''\notin \{a,a'\}$ and $\bm{z}\in \Sigma_q^{\geq 0}$ such that $a''\bm{z} \in \mathcal{S}^{a''}$, then $\bm{z}\in \mathcal{I}_1(\tilde{\bm{x}})\cap \mathcal{I}_1(\tilde{\bm{y}})$, which contradicts the condition that $d_L(\tilde{\bm{x}},\tilde{\bm{y}})\geq 4$.
Similarly, we can obtain $\mathcal{S}_{b''}=\varnothing$ for $b''\not\in \{b,b'\}$.
This implies that $\mathcal{S}= \mathcal{S}_{b}^{a} \sqcup \mathcal{S}_{b}^{a'} \sqcup \mathcal{S}_{ b'}^{a} \sqcup \mathcal{S}_{b'}^{a'}$.
Below, we consider the four sets one by one.

For the set $\mathcal{S}_{b}^{a}$, we have $\mathcal{S}_{b}^{a}=a\circ(\mathcal{I}_{2}(\bm{r}) \cap \{ a'\bm{s}b' \}) \circ b \subseteq \{aa'\bm{s}b'b\}$, then $|\mathcal{S}_{b}^{a}| \leq 1$. Let $\bm{z}\triangleq aa'\bm{s}b'b$. If $|\mathcal{S}_{b}^{a}| =1$, we have $\mathcal{S}_{b}^{a}= \{\bm{z}\}$. Then there exist integers $\ell$ and $k$ with $1 < \ell< k <m+2$, such that $\bm{z}\in \tilde{\bm{x}}(\ell,k)\cap \tilde{\bm{y}}(1,m+2)\subseteq \mathcal{A}_1$. Thus, $\mathcal{S}_{b}^{a} \subseteq \mathcal{A}_1$.
Moreover, by symmetry, we have $\mathcal{S}_{b'}^{a'} \subseteq \mathcal{A}_4$.

For the set $\mathcal{S}_{b'}^{a}$, we have $\mathcal{S}_{b'}^{a}=a \circ (\mathcal{I}_{1}(\bm{r}b) \cap \mathcal{I}_{1}(a'\bm{s})) \circ b'$, it follows by Lemma \ref{lem:position} that $|\mathcal{S}_{b'}^{a}|= |\mathcal{I}_1(\bm{r}b) \cap \mathcal{I}_1(a'\bm{s})| \leq 2$. 
If $|\mathcal{S}_{b'}^{a}|=|\mathcal{I}_1(\bm{r}b) \cap \mathcal{I}_1(a'\bm{s})|=2$, denote $\mathcal{S}_{b'}^{a}= \{\bm{z}, \bm{z}'\}$, again by Lemma \ref{lem:position}, there exist integers $\ell$ and $k$ with $1<\ell< k<m+2$ such that $\bm{z} \in \tilde{\bm{x}}(\ell,m+2)\cap \tilde{\bm{y}}(1,k)\subseteq \mathcal{A}_2$ and $\bm{z}' \in \tilde{\bm{x}}(k,m+2)\cap \tilde{\bm{y}}(1,\ell)\subseteq \mathcal{A}_3$.
Thus, $\mathcal{S}_{b'}^{a} \subseteq \mathcal{A}_2 \cup \mathcal{A}_3$.
Similarly, we can obtain $\mathcal{S}_{b}^{a'} \subseteq \mathcal{A}_5 \cup \mathcal{A}_6$.

In summary, we obtain $\mathcal{S}\subseteq \cup_{i=1}^6 \mathcal{A}_i$. 
By definition, we have $\mathcal{A}_i \subseteq \mathcal{S}$ for $i\in [1,6]$.
This implies that $\mathcal{S}= \cup_{i=1}^6 \mathcal{A}_i$. 
Furthermore, from the discussion above, we can conclude that $|\mathcal{A}_i| \leq 1$ for $i\in [1,6]$.
This completes the proof.
\end{IEEEproof}

Now, we are prepared to characterize pairs of sequences that have a Levenshtein distance of at least four, given that the intersection size of their two-insertion balls is at least $2$, $3$, $4$, and $5$, respectively.

\begin{theorem}\label{thm:characterization}
Let $\bm{x}= \bm{u} \tilde{\bm{x}} \bm{v}, \bm{y}= \bm{u} \tilde{\bm{y}} \bm{v}\in \Sigma_q^n$ be such that $d_L(\bm{x},\bm{y}) \geq 4$.
  \begin{itemize}
    \item If $|\mathcal{I}_2(\bm{x}) \cap \mathcal{I}_2(\bm{y})| \geq 2$, then either $\sum_{i\in \{1,2,4,5\}}|\mathcal{A}_i|\geq 1$ or $\tilde{\bm{x}}$ is a combination of at most five $4^{\leq}$-periodic sequences.
    \item If $|\mathcal{I}_2(\bm{x}) \cap \mathcal{I}_2(\bm{y})| \geq 3$, then $\sum_{i\in \{1,2,4,5\}}|\mathcal{A}_i|\geq 1$.
    \item If $|\mathcal{I}_2(\bm{x}) \cap \mathcal{I}_2(\bm{y})| \geq 4$, then either $\tilde{\bm{x}}= \bm{\alpha}_t(aa') \bm{w} \bm{\alpha}_s(bb')$ and  $\tilde{\bm{y}}= \bm{\alpha}_t(a'a) \bm{w} \bm{\alpha}_s(b'b)$ for some $t,s\geq 1$ or $\tilde{\bm{x}}$ is a combination of at most five $4^{\leq}$-periodic sequences.
    \item If $|\mathcal{I}_2(\bm{x}) \cap \mathcal{I}_2(\bm{y})| \geq 5$, then $\tilde{\bm{x}}$ is a combination of at most five $4^{\leq}$-periodic sequences.
  \end{itemize}
\end{theorem}

\begin{IEEEproof}
    Observe that $\mathcal{I}_2(\bm{x}) \cap \mathcal{I}_2(\bm{y})= \bm{u}\circ (\mathcal{I}_2(\tilde{\bm{x}}) \cap \mathcal{I}_2(\tilde{\bm{y}})) \circ \bm{v}$, then by Lemma \ref{lem:classify}, we have $|\mathcal{I}_2(\bm{x}) \cap \mathcal{I}_2(\bm{y})|= \sum_{i=1}^6 |\mathcal{A}_i|$, where $|\mathcal{A}_i|\leq 1$.
    \begin{itemize}
    \item If $|\mathcal{I}_2(\bm{x}) \cap \mathcal{I}_2(\bm{y})| \geq 2$, we have either $\sum_{i\in \{1,2,4,5\}}|\mathcal{A}_i|\geq 1$ or $|\mathcal{A}_3|=|\mathcal{A}_6|=1$. 
    \item If $|\mathcal{I}_2(\bm{x}) \cap \mathcal{I}_2(\bm{y})| \geq 3$, then $\sum_{i\in \{1,2,4,5\}}|\mathcal{A}_i|\geq 1$.
    \item If $\sum_{i=1}^6 |\mathcal{A}_i|\geq 4$, we have either $|\mathcal{A}_1|=|\mathcal{A}_4|=1$ or $|\mathcal{A}_2|=|\mathcal{A}_6|=1$ or $|\mathcal{A}_3|=|\mathcal{A}_5|=1$.
    \item If $\sum_{i=1}^6 |\mathcal{A}_i|\geq 5$, we have either $|\mathcal{A}_2|=|\mathcal{A}_6|=1$ or $|\mathcal{A}_3|=|\mathcal{A}_5|=1$.
  \end{itemize}
  Then the conclusion follows by Claims \ref{cla:14} and \ref{cla:36}.
\end{IEEEproof}

\begin{claim}\label{cla:14}
    For two distinct sequences $\tilde{\bm{x}}= a \bm{r} b, \tilde{\bm{y}}= a' \bm{s} b'\in \Sigma_q^m$, where $m\geq 2$ and $a, a', b, b'\in \Sigma_q$ with $a\neq a'$ and $b \neq b'$, if $d_L(\tilde{\bm{x}},\tilde{\bm{y}}) \geq 4$ and $|\mathcal{A}_1|=|\mathcal{A}_4|=1$, then we can express $\tilde{\bm{x}}= \bm{\alpha}_t(aa') \bm{w} \bm{\alpha}_s(bb')$ and $\tilde{\bm{y}}= \bm{\alpha}_t(a'a) \bm{w} \bm{\alpha}_s(b'b)$ for some $t,s\geq 1$ and $\bm{w}\in \Sigma_q^{\geq 0}$.
\end{claim}

\begin{IEEEproof}
    When $|\mathcal{A}_1|=1$, there exists some $1\leq \ell_1<k_1\leq m$ such that $\tilde{\bm{x}}(\ell_1+1,k_1+1)\cap \tilde{\bm{y}}(1,m+2)\neq \varnothing$.
    It follows by Table \ref{tab:char} that $\tilde{\bm{x}}_{[1,m]\setminus \{1,m\}}=\tilde{\bm{y}}_{[1,m]\setminus \{\ell_1,k_1\}}$.
    By symmetry, when $|\mathcal{A}_4|=1$, there exists some $1\leq \ell_2<k_2\leq m$ such that $\tilde{\bm{x}}_{[1,m]\setminus \{\ell_2,k_2\}}=\tilde{\bm{y}}_{[1,m]\setminus \{1,m\}}$.
    Then the conclusion follows by \cite[Lemma 3.8 and Figure 1]{Sun-23-IT}.
\end{IEEEproof}

\begin{remark}
    In Claim \ref{cla:14}, we permit the cases where $t=1$ or $s=1$, whereas in \cite[Lemma 3.8 and Figure 1]{Sun-23-IT}, it is assumed that $t, s \geq 2$. This difference arises from the fact that we allow $\ell_i=1$ or $k_i=m$ for some $i \in \{1, 2\}$.
\end{remark}

\begin{claim}\label{cla:36}
    For two distinct sequences $\tilde{\bm{x}}= a \bm{r} b, \tilde{\bm{y}}= a' \bm{s} b'\in \Sigma_q^m$, where $m\geq 2$ and $a, a', b, b'\in \Sigma_q$ with $a\neq a'$ and $b \neq b'$, if $d_L(\tilde{\bm{x}},\tilde{\bm{y}}) \geq 4$ and we have either $|\mathcal{A}_3|=|\mathcal{A}_6|=1$ or $|\mathcal{A}_2|=|\mathcal{A}_6|=1$ or $|\mathcal{A}_3|=|\mathcal{A}_5|=1$, then $\tilde{\bm{x}}$ is a combination of at most five $4^{\leq}$-periodic sequences.
\end{claim}

\begin{table*}
\renewcommand\arraystretch{1.5}
\caption{The structures of $\tilde{\bm{x}}$ and $\tilde{\bm{y}}$ when $|\mathcal{A}_i|=1$ for some $i$.
We assume that $1<\ell_1<k_1<m+2$ and $1<\ell_2<k_2<m+2$}\label{tab:char}
  \centering
  \begin{tabular}{c|c|c}
    \hline
    $|\mathcal{A}_1|=1$ & $\tilde{\bm{x}}(\ell_1,k_1)\cap \tilde{\bm{y}}(1,m+2)\neq \varnothing$ & $\tilde{\bm x}_{[2,\ell_1-1]}=\tilde{\bm y}_{[1,\ell_1-2]},~ \tilde{\bm x}_{[\ell_1,k_1-2]}=\tilde{\bm y}_{[\ell_1,k_1-2]},~ \tilde{\bm x}_{[k_1-1,m-1]}=\tilde{\bm y}_{[k_1,m]}$ \\
    \hline
    $|\mathcal{A}_2|=1$ & $\tilde{\bm{x}}(\ell_1,m+2)\cap \tilde{\bm{y}}(1,k_1)\neq \varnothing$ & $\tilde{\bm x}_{[2,\ell_1-1]}=\tilde{\bm y}_{[1,\ell_1-2]},~ \tilde{\bm x}_{[\ell_1,k_1-2]}=\tilde{\bm y}_{[\ell_1,k_1-2]},~ \tilde{\bm x}_{[k_1,m]}=\tilde{\bm y}_{[k_1-1,m-1]}$ \\
    \hline
    $|\mathcal{A}_3|=1$ & $\tilde{\bm{x}}(k_1,m+2)\cap \tilde{\bm{y}}(1,\ell_1)\neq \varnothing$ & $\tilde{\bm x}_{[2,\ell_1-1]}=\tilde{\bm y}_{[1,\ell_1-2]},~ \tilde{\bm x}_{[\ell_1+1,k_1-1]}=\tilde{\bm y}_{[\ell_1-1,k_1-3]},~ \tilde{\bm x}_{[k_1,m]}=\tilde{\bm y}_{[k_1-1,m-1]}$ \\
    \hline
    $|\mathcal{A}_4|=1$ & $\tilde{\bm{x}}(1,m+2)\cap \tilde{\bm{y}}(\ell_2,k_2)\neq \varnothing$ & $\tilde{\bm x}_{[1,\ell_2-2]}=\tilde{\bm y}_{[2,\ell_2-1]},~ \tilde{\bm x}_{[\ell_2,k_2-2]}=\tilde{\bm y}_{[\ell_2,k_2-2]},~ \tilde{\bm x}_{[k_2,m]}=\tilde{\bm y}_{[k_2-1,m-1]}$ \\
    \hline
    $|\mathcal{A}_5|=1$ & $\tilde{\bm{x}}(1,m+2)\cap \tilde{\bm{y}}(\ell_2,k_2)\neq \varnothing$ & $\tilde{\bm x}_{[1,\ell_2-2]}=\tilde{\bm y}_{[2,\ell_2-1]},~ \tilde{\bm x}_{[\ell_2,k_2-2]}=\tilde{\bm y}_{[\ell_2,k_2-2]},~ \tilde{\bm x}_{[k_2-1,m-1]}=\tilde{\bm y}_{[k_2,m]}$ \\
    \hline
    $|\mathcal{A}_6|=1$ & $\tilde{\bm{x}}(1,\ell_2)\cap \tilde{\bm{y}}(k_2,m+2)\neq \varnothing$ & $\tilde{\bm x}_{[1,\ell_2-2]}=\tilde{\bm y}_{[2,\ell_2-1]},~ \tilde{\bm x}_{[\ell_2-1,k_2-3]}=\tilde{\bm y}_{[\ell_2+1,k_2-1]},~ \tilde{\bm x}_{[k_2-1,m-1]}=\tilde{\bm y}_{[k_2,m]}$ \\
    \hline
  \end{tabular}
\end{table*}

\begin{IEEEproof}
    We first characterize the sequences $\tilde{\bm{x}}$ and $\tilde{\bm{y}}$ when $|\mathcal{A}_i|=1$, for some $i$, and list their structures in Table \ref{tab:char}.
    In what follows, we only consider the case where $|\mathcal{A}_3|=|\mathcal{A}_6|=1$, as the other case can be discussed similarly.
    By symmetry, without loss of generality, we assume that $\ell_1\leq \ell_2$.
    We distinguish between three cases.
    \begin{itemize}
    \item If $\ell_1 \leq \ell_2 \leq k_2 \leq k_1$, we have
        \begin{equation*}
        \begin{cases}
        \tilde{\bm{x}}_{[2,\ell_1-1]}= \tilde{\bm{y}}_{[1,\ell_1-2]} \text{ and } \tilde{\bm{x}}_{[1,\ell_1-2]}= \tilde{\bm{y}}_{[2,\ell_1-1]}, \\
        \tilde{\bm{x}}_{[\ell_1+1,\ell_2-1]}= \tilde{\bm{y}}_{[\ell_1-1,\ell_2-3]} \text{ and } \tilde{\bm{x}}_{[\ell_1-1,\ell_2-2]}= \tilde{\bm{y}}_{[\ell_1,\ell_2-1]}, \\
        \tilde{\bm{x}}_{[\ell_2,k_2-1]}= \tilde{\bm{y}}_{[\ell_2-2,k_2-3]} \text{ and } \tilde{\bm{x}}_{[\ell_2-1,k_2-3]}= \tilde{\bm{y}}_{[\ell_2+1,k_2-1]}, \\
        \tilde{\bm{x}}_{[k_2,k_1-1]}= \tilde{\bm{y}}_{[k_2-2,k_1-3]} \text{ and } \tilde{\bm{x}}_{[k_2-1,k_1-2]}= \tilde{\bm{y}}_{[k_2,k_1-1]}, \\
        \tilde{\bm{x}}_{[k_1,m]}= \tilde{\bm{y}}_{[k_1-1,m-1]} \text{ and } \tilde{\bm{x}}_{[k_1-1,m-1]}= \tilde{\bm{y}}_{[k_1,m]}.
        \end{cases}
        \end{equation*}
        In this case, we can conclude that $\tilde{\bm{x}}_{[1,\ell_1-1]}$ is $2$-periodic, 
        $\tilde{\bm{x}}_{[\ell_1-1,\ell_2-1]}$ is $3$-periodic, $\tilde{\bm{x}}_{[\ell_2-1,k_2-1]}$ is $4$-periodic,  $\tilde{\bm{x}}_{[k_2-1,k_1-1]}$ is $3$-periodic, and
        $\tilde{\bm{x}}_{[k_1-1,m]}$ is $2$-periodic.
    \item If $\ell_1 \leq \ell_2 \leq k_1 \leq k_2$, we have
        \begin{equation*}
        \begin{cases}
        \tilde{\bm{x}}_{[2,\ell_1-1]}= \tilde{\bm{y}}_{[1,\ell_1-2]} \text{ and } \tilde{\bm{x}}_{[1,\ell_1-2]}= \tilde{\bm{y}}_{[2,\ell_1-1]}, \\
        \tilde{\bm{x}}_{[\ell_1+1,\ell_2-1]}= \tilde{\bm{y}}_{[\ell_1-1,\ell_2-3]} \text{ and } \tilde{\bm{x}}_{[\ell_1-1,\ell_2-2]}= \tilde{\bm{y}}_{[\ell_1,\ell_2-1]}, \\
        \tilde{\bm{x}}_{[\ell_2,k_1-1]}= \tilde{\bm{y}}_{[\ell_2-2,k_1-3]} \text{ and } \tilde{\bm{x}}_{[\ell_2-1,k_1-3]}= \tilde{\bm{y}}_{[\ell_2+1,k_1-1]}, \\
        \tilde{\bm{x}}_{[k_1,k_2-1]}= \tilde{\bm{y}}_{[k_1-1,k_2-2]} \text{ and } \tilde{\bm{x}}_{[k_1-2,k_2-3]}= \tilde{\bm{y}}_{[k_1,k_2-1]}, \\
        \tilde{\bm{x}}_{[k_2,m]}= \tilde{\bm{y}}_{[k_2-1,m-1]} \text{ and } \tilde{\bm{x}}_{[k_2-1,m-1]}= \tilde{\bm{y}}_{[k_2,m]}.
        \end{cases}
        \end{equation*}
        In this case, we can conclude that $\tilde{\bm{x}}_{[1,\ell_1-1]}$ is $2$-periodic, 
        $\tilde{\bm{x}}_{[\ell_1-1,\ell_2-1]}$ is $3$-periodic, $\tilde{\bm{x}}_{[\ell_2-1,k_1-1]}$ is $4$-periodic,  $\tilde{\bm{x}}_{[k_1-1,k_2-1]}$ is $3$-periodic, and
        $\tilde{\bm{x}}_{[k_2-1,m]}$ is $2$-periodic.
    \item If $\ell_1 \leq k_1 \leq \ell_2 \leq k_2$, we have
        \begin{equation*}
        \begin{cases}
        \tilde{\bm{x}}_{[2,\ell_1-1]}= \tilde{\bm{y}}_{[1,\ell_1-2]} \text{ and } \tilde{\bm{x}}_{[1,\ell_1-2]}= \tilde{\bm{y}}_{[2,\ell_1-1]}, \\
        \tilde{\bm{x}}_{[\ell_1+1,k_1-1]}= \tilde{\bm{y}}_{[\ell_1-1,k_1-3]} \text{ and } \tilde{\bm{x}}_{[\ell_1-1,k_1-2]}= \tilde{\bm{y}}_{[\ell_1,k_1-1]}, \\
        \tilde{\bm{x}}_{[k_1,\ell_2-1]}= \tilde{\bm{y}}_{[k_1-1,\ell_2-2]} \text{ and } \tilde{\bm{x}}_{[k_1-1,\ell_2-2]}= \tilde{\bm{y}}_{[k_1,\ell_2-1]}, \\
        \tilde{\bm{x}}_{[\ell_2,k_2-1]}= \tilde{\bm{y}}_{[\ell_2-1,k_2-2]} \text{ and } \tilde{\bm{x}}_{[\ell_2-1,k_2-3]}= \tilde{\bm{y}}_{[\ell_2+1,k_2-1]}, \\
        \tilde{\bm{x}}_{[k_2,m]}= \tilde{\bm{y}}_{[k_2-1,m-1]} \text{ and } \tilde{\bm{x}}_{[k_2-1,m-1]}= \tilde{\bm{y}}_{[k_2,m]}.
        \end{cases}
        \end{equation*}
        In this case, we can conclude that $\tilde{\bm{x}}_{[1,\ell_1-1]}$ is $2$-periodic, 
        $\tilde{\bm{x}}_{[\ell_1-1,k_1-1]}$ is $3$-periodic, $\tilde{\bm{x}}_{[k_1-1,\ell_2-1]}$ is $2$-periodic,  $\tilde{\bm{x}}_{[\ell_2-1,k_2-1]}$ is $3$-periodic, and
        $\tilde{\bm{x}}_{[k_2-1,m]}$ is $2$-periodic.
\end{itemize}
In each case, $\tilde{\bm{x}}$ can be viewed as a combination of at most five $4^{\leq}$-periodic sequences.
This completes the proof.
\end{IEEEproof}

\begin{remark}
  There is a slight difference between the case where $|\mathcal{A}_3| = |\mathcal{A}_6| = 1$ and the case depicted in \cite[Figure 3]{Sun-23-IT}. When $|\mathcal{A}_3|=1$, from Table \ref{tab:char}, we obtain $\tilde{\bm{x}}_{[1,m]\setminus \{1,\ell_1\}} = \tilde{\bm{y}}_{[1,m]\setminus \{k_1-2, m\}}$ for some $1 < \ell_1 < k_1 < m + 2$. It is worth noting that in this case, we may have $\ell_1 > k_1 - 2$, whereas in \cite[Figure 3]{Sun-23-IT} it is always assumed that $\tilde{\bm{x}}_{[1,m]\setminus \{1,\ell_1\}} = \tilde{\bm{y}}_{[1,m]\setminus \{k_1-2, m\}}$ for some $\ell_1 \leq k_1 - 2$.
  This subtle difference arises from the fact that, for two sequences with a Hamming distance of one, the intersection size of their single-insertion balls is two, whereas the intersection size of their single-deletion balls is only one.
\end{remark}

\subsection{Tools}
Before proceeding with the code constructions, we introduce several tools that will be used later.

\subsubsection{Tools From \texorpdfstring{\cite{Sun-25-arXiv}}{Sun-25-arXiv}}
We begin by presenting the tools from \cite{Sun-25-arXiv} that are employed for error correction.

\begin{definition}
  Let $\bm{z}$ be a sequence of length $n$ and $k\geq 0$ be an integer, we define the \emph{$k$-th order VT syndrome} of $\bm{z}$ as $\mathrm{VT}^{k}(\bm{z})= \sum_{i=1}^n i^k z_i$.
\end{definition}

\begin{definition}
Let $\bm{z}$ be a sequence with integer entries, we define its \emph{sign-preserving number}, denoted by $\sigma(\bm{z})$, as the smallest positive integer $k$ such that $\bm{z}$ can be partitioned into $k$ substrings, with the property that within each substring, all non-zero entries share the same sign.
\end{definition} 

\begin{definition}\label{def:g} 
For any $\bm{x} \in \Sigma_q^n$, we define its \emph{differential sequence} as $\bm{d}(\bm{x}) = \big(d(\bm{x})_1, d(\bm{x})_2, \ldots, d(\bm{x})_{n}\big)$ and its \emph{accumulative differential sequence} as $\bm{g}(\bm{x}) = \big(g(\bm{x})_1, g(\bm{x})_2, \ldots,g(\bm{x})_{n}\big)$, where $d(\bm{x})_i = x_i - x_{i-1} \pmod{q}$ and $g(\bm{x})_i = \sum_{j=1}^i d(\bm{x})_j $ for $i \in [1,n]$.  
\end{definition} 

\begin{lemma}\cite[Theorem IV.23]{Sun-25-arXiv}\label{thm:q}
  Let $n_1\geq qn$, for $a\in [0,n_1-1]$, the code
  \begin{align*}
    \mathcal{C}= \big\{\bm{x}\in \Sigma_q^n: \mathrm{VT}^0\big(\bm{g}(\bm{x})\big) \equiv a \pmod{n_1} \big\}
  \end{align*}
  is a $q$-ary single-insertion correcting code. Moreover, when $n_1=qn$, there exists a choice of parameters such that its redundancy is at most $\log n+\log q$ bits.
\end{lemma}

\begin{lemma}\cite[Theorem IV.6]{Sun-25-arXiv}\label{lem:x=y}
  Let $\bm{x}$ and $\bm{y}$ be such that $\sigma\big(\bm{g}(\bm{x})-\bm{g}(\bm{y})\big)\leq m$ and $\mathrm{VT}^k\big(\bm{g}(\bm{x})-\bm{g}(\bm{y})\big)=0$ for $k\in [0,m-1]$, then $\bm{x}=\bm{y}$.
\end{lemma} 

\begin{lemma}\cite[Lemma IV.22]{Sun-25-arXiv}\label{lem:partition}
  Let $\bm{x},\bm{y}\in \Sigma_q^n$ be such that $\bm{x}=\bm{x}^{(1)}\bm{x}^{(2)}\cdots \bm{x}^{(m)}$ and $\bm{y}=\bm{y}^{(1)}\bm{y}^{(2)}\cdots \bm{y}^{(m)}$, where $m\geq 1$ and $|\bm{x}^{(i)}|=|\bm{y}^{(i)}|$ for $i\in [1,m]$.
  If $\mathcal{D}_{1}(\bm{x}^{(i)})\cap \mathcal{D}_{1}(\bm{y}^{(i)})\neq \varnothing$, then $\sigma\big(\bm{g}(\bm{x})-\bm{g}(\bm{y})\big)\leq m$ and for $k\geq 0$, $\big| \mathrm{VT}^k\big(\bm{g}(\bm{x})-\bm{g}(\bm{y})\big) \big|\leq qm\sum_{i=1}^n i^k-m$.
\end{lemma}

Now we use Lemmas \ref{lem:x=y} and \ref{lem:partition} to correct two insertions in $q$-ary alphabets under several conditions, which will be applied in the construction of $(n,q,N;\mathcal{I}_2)$-reconstruction codes for $N\in \{2,3\}$.

\begin{lemma}\label{lem:N=3}
  For $k\in [0,1]$, let $n_k=2q\sum_{i=1}^n i^k-2$ and $a_k\in [0,n_k]$, we define the code
  \begin{align*}
    \mathcal{C}_{1,2,4,5}= \big\{\bm{x}\in \Sigma_q^n: \mathrm{VT}^k\big(\bm{g}(\bm{x})\big) \equiv a_k \pmod{n_k+1} \text{ for } k\in [0,1]\big\}.
  \end{align*}
  Then for any $\bm{x}\neq \bm{y} \in \mathcal{C}_{1,2,4,5}$, we have $d_L(\bm{x},\bm{y})\geq 4$ and $|\mathcal{A}_i|=0$ for $i\in \{1,2,4,5\}$.
\end{lemma}

\begin{IEEEproof}
  Observe that $\mathcal{C}_{1,2,4,5}$ is a subcode of the single-insertion correcting code defined in Lemma \ref{thm:q}, then $d_L(\bm{x},\bm{y})\geq 4$.
  When $|\mathcal{A}_i|\neq 0$ for some $i\in \{1,2,4,5\}$, by Table \ref{tab:char}, we can conclude that there exists a partition 
  \begin{gather*}
  \bm{x} = \bm{x}^{(1)} \bm{x}^{(2)},\quad
  \bm{y} = \bm{y}^{(1)} \bm{y}^{(2)},
  \end{gather*}
  such that for each $ i \in [1, 2] $, $|\bm{x}^{(i)}| = |\bm{y}^{(i)}|$ and $\mathcal{D}_{1}(\bm{x}^{(i)}) \cap \mathcal{D}_{1}(\bm{y}^{(i)}) \neq \varnothing$.
  Then by Lemma \ref{lem:partition}, we have $\sigma(\bm{g}(\bm{x})-\bm{g}(\bm{y}))\leq 2$ and for $k\in \{0,1\}$, $|\mathrm{VT}^{k}(\bm{g}(\bm{x})-\bm{g}(\bm{y}))|\leq 2q\sum_{i=1}^n i^k-2=n_k$.
  Since $\mathrm{VT}^k\big(\bm{g}(\bm{x})\big)\equiv \mathrm{VT}^k\big(\bm{g}(\bm{y})\big) \pmod{n_k+1}$, we get $\mathrm{VT}^{k}(\bm{g}(\bm{x})-\bm{g}(\bm{y}))=0$.
  It follows by Lemma \ref{lem:x=y} that $\bm{x}=\bm{y}$, which leads to a contradiction.
  Thus, we have $|\mathcal{A}_i|=0$ for $i\in \{1,2,4,5\}$.
  This completes the proof.
\end{IEEEproof}

\subsubsection{Tools From \texorpdfstring{\cite{Sun-25-IT-C}}{Sun-25-IT-C}}

We now introduce some tools from \cite{Sun-25-IT-C} that will be employed in the construction of $(n, q, 4; \mathcal{I}_2)$-reconstruction codes.

\begin{lemma}\cite[Theorems 3 and 14]{Sun-25-IT-C}\label{lem:aux}
  Let $P= 3\lceil \log_q n+\log_q\log_q n \rceil+4$ and let $p\geq 4n$ be a prime (note that we can choose $p\in [4n,8n]$).
  Let $d_1 \in [0,2q-2], d_2 \in [0,p-1], d_3 \in [0,\frac{(q-1)(P-1)}{3}]$, we define the code $\mathcal{C}_{N=4}^{aux}\subseteq \Sigma_q^n$ in which each sequence $\bm{x}$ satisfies the following conditions:
  \begin{itemize}
    \item $\mathrm{VT}^0(\bm{x}) \equiv d_1 \pmod{2q-1}$;
    \item $\mathrm{VT}^2(\bm{x}) \equiv d_2 \pmod{p}$;
    \item $\mathrm{VT}^0(\mathcal{O}(\bm{x})) \equiv d_3 \pmod{\frac{(q-1)(P-1)}{3}+1}$, where $\mathcal{O}(\bm{x})=x_1x_3\cdots x_{2\lceil n/2 \rceil-1}$.
  \end{itemize}
  For any $\bm{x}, \bm{y} \in \Sigma_q^n$, if there exists some $\bm{u}, \bm{v},\bm{w} \in \Sigma_q^{\geq 0}, \bm{v}  \in \Sigma_q^{\geq 1}$, $t_1,t_2 \geq 1$, and $a_1, b_1, a_2, b_2 \in \Sigma_q$ with $a_1\neq b_1$ and $a_2\neq b_2$, such that
  \begin{gather*}
      \bm{x}= (\bm{u}, \bm{\alpha}_{t_1}(a_1b_1), \bm{w}, \bm{\alpha}_{t_2}(a_2b_2), \bm{v}); \\
      \bm{y}= (\bm{u}, \bm{\alpha}_{t_1}(b_1a_1), \bm{w}, \bm{\alpha}_{t_2}(b_2a_2), \bm{v}),
  \end{gather*}
  then $\bm{x}$ and $\bm{y}$ can not belong to $\mathcal{C}_{N=4}^{aux}\cap \mathcal{R}_q(n,2,\frac{P-1}{3})$ simultaneously (note that $\mathcal{R}_q(n,4,\frac{P-1}{3})\subseteq \mathcal{R}_q(n,2,\frac{P-1}{3})$).
\end{lemma}

\subsubsection{Tools From \texorpdfstring{\cite{Sun-23-IT}}{Sun-23-IT}}


The following lemma is a corollary of \cite[Theorem 5.6]{Sun-23-IT}.

\begin{lemma}\cite[Theorem 5.6]{Sun-23-IT}\label{lem:P}
    There exists a function $\xi:\Sigma_2^n\rightarrow [0,O(P^{14})]$, computable in linear
time, such that for any two distinct sequences $\bm{x}= \bm{u} \tilde{\bm{x}} \bm{v}, \bm{y}= \bm{u} \tilde{\bm{y}} \bm{v}\in \Sigma_2^n$ with $|\tilde{\bm x}|\leq P$ and $\xi(\bm{x})=\xi(\bm{y})$, it holds that $\mathcal{I}_2(\bm{x}) \cap \mathcal{I}_2(\bm{y})=\varnothing$.
\end{lemma}

In the following, we extend this result to non-binary alphabets.

\begin{definition}
  For any $\bm{x}\in \Sigma_q^n$, we define its \emph{binary matrix representation} as
  \begin{equation*}
    M(\bm{x})=
    \begin{pmatrix}
      x_{1,1} & \cdots & x_{1,n} \\
      \vdots & \ddots & \vdots \\
      x_{\lceil \log q \rceil,1} & \cdots & x_{\lceil \log q \rceil,n} 
    \end{pmatrix},
  \end{equation*}
  where $x_{k,i}\in\{0,1\}$ such that $x_i=\sum_{k=1}^{\lceil \log q \rceil}x_{k,i}2^{k-1}$ for $i\in [1,n]$.
  Let $M_i(\bm{x})$ be the $i$-th row of $M(\bm{x})$ for $i\in [1,\lceil \log q \rceil]$. 
\end{definition}

\begin{lemma}\label{lem:P'}
    Let $\xi(\cdot)$ be defined in Lemma \ref{lem:P}.
    For any two distinct sequences $\bm{x}= \bm{u} \tilde{\bm{x}} \bm{v}, \bm{y}= \bm{u} \tilde{\bm{y}} \bm{v}\in \Sigma_q^n$ with $|\tilde{\bm x}|\leq P$ and $\xi(M_i(\bm{x}))=\xi(M_i(\bm{y}))$ for $i\in [1,\lceil \log q \rceil]$, it holds that $\mathcal{I}_2(\bm{x}) \cap \mathcal{I}_2(\bm{y})=\varnothing$.
\end{lemma}

\begin{IEEEproof}
    Assume that $\mathcal{I}_2(\bm{x}) \cap \mathcal{I}_2(\bm{y})\neq \varnothing$, then there exists some $i\in [1,\lceil \log q\rceil]$ such that $M_i(\bm{x})\neq M_i(\bm{y})$ and $\mathcal{I}_2(M_i(\bm{x})) \cap \mathcal{I}_2(M_i(\bm{y}))\neq \varnothing$.
    However, since $\xi(M_i(\bm{x}))=\xi(M_i(\bm{y}))$, it follows by Lemma \ref{lem:P} that $\mathcal{I}_2(M_i(\bm{x})) \cap \mathcal{I}_2(M_i(\bm{y}))=\varnothing$, which leads to a contradiction.
    Thus, we have $\mathcal{I}_2(\bm{x}) \cap \mathcal{I}_2(\bm{y})=\varnothing$.
    This completes the proof.
\end{IEEEproof}

By Lemma \ref{lem:P'}, we can construct the following code, which will be utilized in the construction of $(n,q,N;\mathcal{I}_2)$-reconstruction codes for $N\in \{2,4,5\}$.

\begin{lemma}\label{lem:bounded}
    Let $\xi(\cdot)$ be defined in Lemma \ref{lem:P}.
    For $a_1,a_2,\ldots,a_{\lceil \log q\rceil}\in [0,O((5P)^{14})]$, we define the code
    \begin{gather*}
      \mathcal{C}_{P}=\{\bm{x}\in \Sigma_q^n: \xi(M_i(\bm{x}))=a_i \text{ for } i\in [1,\lceil \log q\rceil] \}.
    \end{gather*}
    For any two distinct sequences $\bm{x}= \bm{u} \tilde{\bm{x}} \bm{v}, \bm{y}= \bm{u} \tilde{\bm{y}} \bm{v}\in \mathcal{R}_q(n,4,P)\cap \mathcal{C}_{P}$, if $\tilde{\bm{x}}$ is a combination of at most five $4^{\leq }$-periodic sequences, then $\mathcal{I}_2(\bm{x}) \cap \mathcal{I}_2(\bm{y})= \varnothing$.
\end{lemma}

\begin{IEEEproof}
    Since $\tilde{\bm{x}}\in \mathcal{R}_q(n,4,P)$ is a combination of at most five $4^{\leq }$-periodic sequences, we have $|\tilde{\bm{x}}|\leq 5P$.
    Then the conclusion follows by Lemma \ref{lem:P'}.
\end{IEEEproof}

The following lemma provides a lower bound on the size of $\mathcal{R}_q(n, t, P)$, which will be instrumental in evaluating our code size.

\begin{lemma}\cite[Lemma 6.10]{Sun-23-IT-burst}\label{lem:period}
    For any positive $P$, if $P\geq\lceil\log n \rceil+t+1$, then $|\mathcal{R}_q(n,t,P)|\geq\frac{q^n}{2}$.
\end{lemma}

\subsection{Constructions}

Now, we are prepared to present our constructions of reconstruction codes.

\begin{theorem}[$N=7$]\label{thm:N=7}
  Let $\mathcal{C}_{N=7}$ be the single-insertion correcting code defined in Lemma \ref{thm:q}, then $\mathcal{C}_{N=7}$ is an $(n,q,7;\mathcal{I}_2)$-reconstruction code with at most $\log n+\log q$ bits of redundancy.
\end{theorem}

\begin{IEEEproof}
For any two distinct sequences $\bm{x},\bm{y}\in \mathcal{C}_{N=7}$, we have $d_L(\bm{x},\bm{y})\geq 4$.
By Lemma \ref{lem:classify}, we get $|\mathcal{I}_2(\bm{x})\cap \mathcal{I}_2(\bm{y})|\leq 6$.
Then the conclusion follows.
\end{IEEEproof}

\begin{theorem}[$N=5$]\label{thm:N=5}
  Let $\mathcal{C}_{N=7}$ be the single-insertion correcting code defined in Theorem \ref{thm:N=7}, and let $\mathcal{C}_{P}$ be code defined in Lemma \ref{lem:bounded}, then $\mathcal{C}_{N=5}=\mathcal{C}_{N=7}\cap \mathcal{C}_{\frac{P-1}{3}}\cap \mathcal{R}_q(n,4,\frac{P-1}{3})$ is an $(n,q,5;\mathcal{I}_2)$-reconstruction code with at most $\log n+O(\log\log n)$ bits of redundancy when $P=3\lceil \log_q n+\log_q\log_q n \rceil+4$.
\end{theorem}

\begin{IEEEproof}
For any two distinct sequences $\bm{x},\bm{y}\in \mathcal{C}_{N=5}$, since $\mathcal{C}_{N=5} \subseteq \mathcal{C}_{N=7}$, we have $d_L(\bm{x},\bm{y})\geq 4$.
Assume $|\mathcal{I}_2(\bm{x})\cap \mathcal{I}_2(\bm{y})|\geq 5$, by Theorem \ref{thm:characterization}, we can write $\bm{x}= \bm{u} \tilde{\bm{x}} \bm{v}$ and $\bm{y}= \bm{u} \tilde{\bm{y}} \bm{v}$, where $\tilde{\bm{x}}$ is a combination of at most five $4^{\leq}$-periodic sequences.
Since $\mathcal{C}_{N=5}\subseteq \mathcal{C}_{\frac{P-1}{3}}\cap \mathcal{R}_q(n,4,\frac{P-1}{3})$, by Lemma \ref{lem:bounded}, we can conclude that $\mathcal{I}_2(\bm{x})\cap \mathcal{I}_2(\bm{y})= \varnothing$, which leads to a contradiction.
Consequently, $\mathcal{C}_{N=5}$ is an $(n,q,5;\mathcal{I}_2)$-reconstruction code.
Moreover, by pigeon-hole principle, there exists a choice of parameters, such that 
\begin{gather*}
  |\mathcal{C}_{N=5}|\geq \frac{|\mathcal{R}_q(n,4,\frac{P-1}{3})|}{O(nq\cdot (5\frac{P-1}{3})^{14 \lceil \log q \rceil})}= \frac{|\mathcal{R}_q(n,4,\frac{P-1}{3})|}{O(nP^{14 \lceil \log q \rceil})}.
\end{gather*}
By Lemma \ref{lem:period}, when $P=3\lceil \log_q n+\log_q\log_q n \rceil+4$, the corresponding redundancy of $\mathcal{C}_{N=5}$ is at most $\log n+O(\log\log n)$ bits.
\end{IEEEproof}

\begin{theorem}[$N=4$]\label{thm:N=4}
  Let $\mathcal{C}_{N=5}$ be the $(n,q,5;\mathcal{I}_2)$-reconstruction code defined in Theorem \ref{thm:N=5}, and let $\mathcal{C}_{N=4}^{aux}$ be the code defined in Lemma \ref{lem:aux}, then the code $\mathcal{C}_{N=4}= \mathcal{C}_{N=5}\cap \mathcal{C}_{N=4}^{aux}$ is an $(n,q,4;\mathcal{I}_2)$-reconstruction code with at most $2\log n+O(\log\log n)$ bits of redundancy when $P= 3\lceil \log_q n+\log_q\log_q n \rceil+4$.
\end{theorem}

\begin{IEEEproof}
    For any two distinct sequences $\bm{x}=\bm{u}\tilde{\bm{x}}\bm{v},\bm{y}=\bm{u}\tilde{\bm{y}}\bm{v}\in \mathcal{C}_{N=4}$, since $\mathcal{C}_{N=4}\subseteq \mathcal{C}_{N=7}$, we have $d_L(\bm{x},\bm{y})\geq 4$. Assume $|\mathcal{I}_2(\bm{x})\cap \mathcal{I}_2(\bm{y})|\geq 4$, since $\mathcal{C}_{N=4} \subseteq \mathcal{C}_{N=5}$, by Theorem \ref{thm:characterization}, it suffices to consider $\tilde{\bm{x}}= \bm{\alpha}_t(aa') \bm{w} \bm{\alpha}_s(bb')$ and  $\tilde{\bm{y}}= \bm{\alpha}_t(a'a) \bm{w} \bm{\alpha}_s(b'b)$ for some $t,s\geq 1$ and $\bm{w}\in \Sigma_q^{\geq 0}$. However, since $\mathcal{C}_{N=4}\subseteq \mathcal{C}_{N=4}^{aux}\cap \mathcal{R}_q(n,4,\frac{P-1}{3})$, this contradicts the conclusion of Lemma \ref{lem:aux}. 
    Consequently, $\mathcal{C}_{N=4}$ is an $(n,q,4;\mathcal{I}_2)$-reconstruction code. 
    Moreover, by pigeon-hole principle, there exists a choice of parameters, such that 
    \begin{gather*}
      |\mathcal{C}_{N=4}|\geq \frac{|\mathcal{R}_q(n,4,P)|}{O(nP^{14 \lceil \log q\rceil})\times O(nP)}= \frac{|\mathcal{R}_q(n,4,P)|}{O(n^2 P^{14\lceil \log q\rceil+1})}.
    \end{gather*}
    By Lemma \ref{lem:period}, when $P=3\lceil \log_q n+\log_q\log_q n \rceil+4$, the corresponding redundancy of $\mathcal{C}_{N=4}$ is at most $2\log n+O(\log\log n)$ bits.
\end{IEEEproof}

\begin{theorem}[$N=3$]\label{thm:N=3}
  Let $\mathcal{C}_{N=3}=\mathcal{C}_{1,2,4,5}$ be the code defined in Lemma \ref{lem:N=3}, then $\mathcal{C}_{N=3}$ is an $(n,q,3;\mathcal{I}_2)$-reconstruction code with at most $3\log n+O(1)$ bits of redundancy.
\end{theorem}

\begin{IEEEproof}
    Since $\mathcal{C}_{N=3}=\mathcal{C}_{1,2,4,5}$, by Lemma \ref{lem:N=3}, for any $\bm{x}\neq \bm{y} \in \mathcal{C}_{N=3}$, we have $d_L(\bm{x},\bm{y})\geq 4$ and $|\mathcal{A}_i|=0$ for $i\in \{1,2,4,5\}$.
    Assume $|\mathcal{I}_2(\bm{x})\cap \mathcal{I}_2(\bm{y})|\geq 3$, by Theorem \ref{thm:characterization}, we have $\sum_{i\in \{1,2,4,5\}} |\mathcal{A}_i|\geq 1$, which leads to a contradiction.
    Therefore, $\mathcal{C}_{N=3}$ is an $(n,q,3;\mathcal{I}_2)$-reconstruction code.
    Moreover, by pigeon-hole principle, there exists a choice of parameters, such that 
    \begin{gather*}
      |\mathcal{C}_{N=3}|\geq \frac{q^n}{\prod_{k=0}^1 (2q-\sum_{i=1}^n i^k-2)}= \frac{q^n}{O(n^3)},
    \end{gather*}
    or equivalently, the corresponding redundancy of $\mathcal{C}_{N=3}$ is at most $3\log n+O(1)$ bits.
\end{IEEEproof}

\begin{theorem}[$N=2$]\label{thm:N=2}
  Let $\mathcal{C}_{N=3}$ be the code defined in Theorem \ref{thm:N=3}, and let $\mathcal{C}_{P}$ be the code defined in Lemma \ref{lem:bounded}, then $\mathcal{C}_{N=2}=\mathcal{C}_{N=3}\cap \mathcal{C}_P\cap \mathcal{R}_q(n,4,P)$ is an $(n,q,2;\mathcal{I}_2)$-reconstruction code with at most $3\log n+O(\log\log n)$ bits of redundancy when $P=\lceil\log n \rceil+5$.
\end{theorem}

\begin{IEEEproof}
    Since $\mathcal{C}_{N=2}\subseteq \mathcal{C}_{N=3}$, for any $\bm{x}\neq \bm{y} \in \mathcal{C}_{N=3}$, we have $d_L(\bm{x},\bm{y})\geq 4$ and $|\mathcal{A}_i|=0$ for $i\in \{1,2,4,5\}$.
    Assume $|\mathcal{I}_2(\bm{x})\cap \mathcal{I}_2(\bm{y})|\geq 3$, by Theorem \ref{thm:characterization}, it suffices to consider $\bm{x}= \bm{u} \tilde{\bm{x}} \bm{v}$ and $\bm{y}= \bm{u} \tilde{\bm{y}} \bm{v}$, where $\tilde{\bm{x}}$ is a combination of at most five $4^{\leq}$-periodic sequences.
    Since $\mathcal{C}_{N=2}\subseteq \mathcal{C}_P\cap \mathcal{R}_q(n,4,P)$, by Lemma \ref{lem:bounded}, we can conclude that $\mathcal{I}_2(\bm{x})\cap \mathcal{I}_2(\bm{y})= \varnothing$, which leads to a contradiction.
    Consequently, $\mathcal{C}_{N=2}$ is an $(n,q,2;\mathcal{I}_2)$-reconstruction code.
    Moreover, by pigeon-hole principle, there exists a choice of parameters, such that 
    \begin{gather*}
      |\mathcal{C}_{N=2}|\geq \frac{|\mathcal{R}_q(n,4,P)|}{O(n^3)\times O(P^{14\lceil \log q\rceil})}= \frac{|\mathcal{R}_q(n,4,P)|}{O(n^3 P^{14\lceil \log q\rceil})}.
    \end{gather*}
    By Lemma \ref{lem:period}, when $P=\lceil\log n \rceil+5$, the corresponding redundancy of $\mathcal{C}_{N=2}$ is at most $3\log n+O(\log\log n)$ bits.
\end{IEEEproof}

\bibliographystyle{IEEEtran}
\bibliography{ref}
\end{document}